\documentclass[12pt]{article}

\pdfoutput=1
\DeclareFontFamily{T1}{calligra}{}
\DeclareFontShape{T1}{calligra}{m}{n}{<->s*[1.44]callig15}{}
\DeclareMathAlphabet\mathcalligra   {T1}{calligra} {m} {n}
\DeclareMathAlphabet\mathzapf       {T1}{pzc} {mb} {it}
\DeclareMathAlphabet\mathchorus     {T1}{qzc} {m} {n}
\DeclareMathAlphabet\mathrsfso      {U}{rsfso}{m}{n}
\DeclareMathAlphabet\mathfrcal      {T1}{frcursive}{m}{it}
\DeclareFontFamily{T1}{frcursive}{}
\DeclareFontShape{T1}{frcursive}{m}{n}{<->s*[1.44]callig15}{}
\DeclareMathAlphabet\mathfrcal      {T1}{frcursive}{m}{it}

\usepackage{amsmath}
\usepackage{amssymb}
\usepackage{graphicx}
\usepackage[dvipsnames]{xcolor}
\usepackage{soul}
\numberwithin{equation}{section}
\usepackage{array}
\usepackage{mathtools}
\usepackage{dsfont}
\usepackage{mathrsfs}

\usepackage{BOONDOX-uprscr}

\usepackage{tikz}\usetikzlibrary{matrix,fit}
\usepackage{tikz-cd} 
\usepackage{varwidth}
\usepackage{enumerate}
\usepackage{appendix}
\usepackage{xfrac}
\usepackage{nicefrac}
\usepackage{mathtools,slashed}

\usepackage[margin=1in]{geometry}
\usepackage{nicematrix}

\usepackage[
    backend=bibtex,
    style=alphabetic,
maxbibnames=99,
giveninits=true,
minalphanames=1,
maxalphanames=3,url=false
  ]{biblatex}

\bibliography{SUSYGrassmannian}

\usepackage{mathabx}
\usepackage{empheq}

\usepackage{setspace}

\usepackage{slashed}
\usepackage{upgreek}
\usepackage{appendix}

\usepackage{tabstackengine}

\usepackage{wrapfig}
\usepackage[abs]{overpic}

\usepackage{float}

\fixTABwidth{T}

\newdimen\mytextwidth
\newcommand\rem[2][cyan!40!green]{\noindent\nobreak\hfil\penalty1000\hfilneg
\mytextwidth=\linewidth\advance\mytextwidth by 2mm
\begin{tikzpicture}[baseline=-\the\dimexpr\fontdimen22\textfont2\relax]\node[outer sep=0pt,draw=black,fill=#1,fill opacity=1,text opacity=1,rectangle,rounded corners]{\begin{varwidth}{\mytextwidth}\textcolor{white}{#2}\end{varwidth}};
\end{tikzpicture}\allowbreak
}

\newcommand\whiterem[2][white!]{\noindent\nobreak\hfil\penalty1000\hfilneg
\mytextwidth=\linewidth\advance\mytextwidth by 2mm
\begin{tikzpicture}[baseline=-\the\dimexpr\fontdimen22\textfont2\relax]\node[outer sep=0pt,draw=black,fill=#1,fill opacity=1,text opacity=1,rectangle,rounded corners,line width=1.5pt]{\begin{varwidth}{\mytextwidth}\textcolor{black}{#2}\end{varwidth}};
\end{tikzpicture}\allowbreak
}

\makeatletter
\newsavebox{\@brx}
\newcommand{\llangle}[1][]{\savebox{\@brx}{\(\m@th{#1\langle}\)}%
  \mathopen{\copy\@brx\kern-0.5\wd\@brx\usebox{\@brx}}}
\newcommand{\rrangle}[1][]{\savebox{\@brx}{\(\m@th{#1\rangle}\)}%
  \mathclose{\copy\@brx\kern-0.5\wd\@brx\usebox{\@brx}}}
\makeatother

\newcommand{\dd}{\partial}

\renewcommand{\tilde}{\widetilde}

\newcommand{\bea}{\begin{equation}}
\newcommand{\eea}{\end{equation}}
\newcommand{\bear}{\begin{eqnarray}}
\newcommand{\eear}{\end{eqnarray}}
\newcommand{\bearr}{\begin{eqnarray*}}
\newcommand{\eearr}{\end{eqnarray*}}

\newcommand{\dif}{\mathrm{d}}

\usepackage{ytableau}
\ytableausetup{centertableaux}

\usepackage{tikz}
\usepackage{xparse}
\NewDocumentCommand{\xrightarrows}{ O{}O{} }{%
\mathrel{%
\vcenter{\hbox{%
\begin{tikzpicture}
  \node[minimum width=1cm,minimum height=1ex,anchor=south,align=center] (a){\text{\vphantom{hg}#1}\\[0.5ex] \vphantom{hg}#2};
  \draw[<-] ([yshift=0.35ex]a.west) -- ([yshift=0.35ex]a.east);
  \draw[->] ([yshift=-0.35ex]a.west) -- ([yshift=-0.35ex]a.east);
\end{tikzpicture}
}}%
}%
}

\renewbibmacro{in:}{}

\usepackage{mdframed}

\newbibmacro{string+doi}[1]{
  \iffieldundef{doi}{#1}{\href{http://dx.doi.org/\thefield{doi}}{#1}}}
\DeclareFieldFormat{title}{\usebibmacro{string+doi}{\mkbibemph{#1}}}
\DeclareFieldFormat[article]{title}{\usebibmacro{string+doi}{\mkbibquote{#1}}}

\newmdenv[
  topline=false,
  bottomline=false,
  rightline=false,
  linewidth=2pt,
  skipabove=\topsep,
  skipbelow=\topsep
]{siderules}

\newmdenv[
  topline=false,
  bottomline=false,
  linewidth=2pt,
  skipabove=\topsep,
  skipbelow=\topsep
]{siderulesright}

\makeatletter
\renewcommand{\@seccntformat}[1]{\csname the#1\endcsname.\quad}
\makeatother

\usepackage{setspace}
\usepackage{xpatch}

\makeatletter
\renewcommand{\@chap@pppage}{
  \clear@ppage
  \thispagestyle{plain}
  \if@twocolumn\onecolumn\@tempswatrue\else\@tempswafalse\fi
  \null\vfil
  \markboth{}{}
  {\centering
   \interlinepenalty \@M
   \normalfont
   \MakeUppercase \appendixpagename\par}
  \if@dotoc@pp
    \addappheadtotoc
  \fi
  \vfil\newpage
  \if@twoside
    \if@openright
      \null
      \thispagestyle{empty}
      \newpage
    \fi
  \fi
  \if@tempswa
    \twocolumn
  \fi
}
\makeatother

\definecolor{navycol}{RGB}{100,150,160}
   \definecolor{pinkcol}{RGB}{242,55,55}
   \definecolor{greencol}{RGB}{50,205,50}

   \definecolor{bluecol}{RGB}{30,144,255}

\usepackage{titlesec}

\titleformat*{\section}{\large\bfseries}
\titleformat*{\subsection}{\normalsize\bfseries}
\titleformat*{\subsubsection}{\normalsize\bfseries}
\titleformat*{\paragraph}{\large\bfseries}
\titleformat*{\subparagraph}{\large\bfseries}
\titlespacing{\author}{-5pt}{-5pt}{-5pt}[-5pt]

\makeatletter
\renewcommand\subsubsection{\@startsection{subsubsection}{3}{\z@}
                                     {-3.25ex\@plus -1ex \@minus -.2ex}
                                     {-1.5ex \@plus -.2ex}
                                     {\normalfont\normalsize\bfseries}}
\renewcommand\subsection{\@startsection{subsection}{3}{\z@}
                                     {-3.25ex\@plus -1ex \@minus -.2ex}
                                     {-1.5ex \@plus -.2ex}
                                     {\normalfont\normalsize\bfseries}}                                     
\makeatother

\DeclareFontFamily{U}{solomos}{}
\DeclareErrorFont{U}{solomos}{m}{n}{10}
\DeclareFontShape{U}{solomos}{m}{n}{
  <-> s*[1.1]  gsolomos8r
}{}

   \usepackage{stmaryrd}

\usepackage{tikz}
\usetikzlibrary{arrows.meta}

\usepackage{indentfirst}

\usepackage{tocloft}
\let \savenumberline \numberline
\def \numberline#1{\savenumberline{#1.}}

\usepackage{etoolbox}
\patchcmd{\tableofcontents}{\@starttoc}{\vspace{-0.3cm}\@starttoc}{}{}

\usepackage{accents}

\newcommand\smallthickbar[1]{\accentset{\rule{.5em}{.7pt}}{#1}}
\usepackage{hyperref}
\hypersetup{
colorlinks=true,
linkcolor=MidnightBlue,
citecolor=violet,
filecolor=purple,
urlcolor=cyan,
breaklinks=true
}

\newcounter{Chapcounter}

\newcommand{\chapter}[1] 
{ {\centering          
  \addtocounter{Chapcounter}{1} \Large \underline{\sffamily \texorpdfstring{\textbf{  Chapter \theChapcounter: ~#1}}{Lg}} }   
  \addcontentsline{toc}{section}{ \color{MidnightBlue} \texorpdfstring{Chapter ~}{Lg}\theChapcounter.\texorpdfstring{~~}{Lg} #1 }    
}

\title{\vspace{-1.0cm} \textbf{Holomorphic Quantization  \\ in Constant Curvature Backgrounds}
\vspace{0.5cm}
}

\author{
 Dmitri Bykov$^{\,a,\,b,\,c,\,d,\,e}$\footnote{Emails:
 bykov@mi-ras.ru, dmitri.v.bykov@gmail.com} \quad  Viacheslav Krivorol$^{\,b,\,c}$\footnote{Email:
 v.a.krivorol@gmail.com}
\\  \vspace{0cm}  \\
{\small $a)$ 
\emph{Steklov
Mathematical Institute of Russian Academy of Sciences,}} \\{\small \emph{Gubkina str. 8, 119991 Moscow, Russia} }\\[4pt]
{\small $b)$ 
\emph{Institute for Theoretical and Mathematical Physics,}} \\{\small \emph{Lomonosov Moscow State University, 119991 Moscow, Russia}}\\[4pt]
{\small $c)$ \emph{HSE University, 6 Usacheva str., Moscow 119048, Russia}}\\[4pt]
{\small $d)$ \emph{Moscow Center of Fundamental and Applied Mathematics,}} \\{\small \emph{ Lomonosov Moscow State University, 119991 Moscow, Russia}}\\[4pt]
{\small $e)$ \emph{Beijing Institute of Mathematical Sciences and Applications (BIMSA),}} \\{\small \emph{Huairou District, Beijing
101408, China}}
}

\date{}

\begin{document}

\maketitle

\vspace{0cm}
\textbf{Abstract.} 
We present a holomorphic quantization scheme for free point particles on two-dimensional constant curvature Riemannian  backgrounds. The procedure is based on a Lagrangian embedding of the particle configuration space into a product of  coadjoint orbits of the background isometry group. Examples are provided by particles on the plane, torus, sphere, and hyperbolic plane, with or without a monopole field. We elaborate the method by recovering the Hamiltonian spectrum and the wave functions on such spaces. As a by-product, we obtain a geometric and physical interpretation of Repka's result on the decomposition of tensor products of $\mathbf{SL}(2,\mathbb{R})$ discrete series representations.

\newpage
\tableofcontents

\newpage

\section{Introduction}
Classical and quantum particles moving in nontrivial geometric backgrounds, or $1D$ sigma models, are of considerable interest from the standpoint of quantum field theory, gravity, and condensed matter theory. In this work, we focus on the simplest possible nontrivial geometries. Specifically, we consider free point particles on two-dimensional, orientable, constant curvature Riemannian manifolds in a transverse magnetic field. 
There exist only three such simply-connected model geometries: the complex plane~$\mathbb{C}$ (zero curvature), the sphere~$\mathbb{S}^2$ (positive curvature), and the Lobachevsky hyperbolic plane~$\mathbb{H}$ (negative curvature), which can also serve as universal covers for tori and hyperbolic surfaces of genus greater than two. A point particle on such manifolds in a perpendicular magnetic field~\cite{DUNNE1992233} not only serves as a good toy model for studying the influence of curvature and topology on the behavior of quantum systems but is also a direct subject of interest from the perspective of the quantum Hall effect \cite{Haldane:1983xm,Iengo:1993cs, Klevtsov:2017fwn, Klevtsov:2015nca, Carlos, Hatsugai}. The sphere and the Lobachevsky plane also represent close Euclidean analogues of the $\mathrm{dS}$ and $\mathrm{AdS}$ geometries, respectively, where the study of particles is interesting in the context of holography and string theory \cite{Dorn:2005jt,Dorn:2005ja,Heinze:2015oha}.

Of course, the subject has been extensively studied. The quantization of a particle on a sphere in a magnetic monopole background originates in the seminal works \cite{Dirac:1931kp,Tamm:1931dda,WU1976365}. The wavefunctions of a charged particle on tori in a magnetic field were analyzed in~\cite{Novikov1, novikov1981magnetic}; see also the treatment in \cite{Dereli:2021wqj,OnofriTorus,Fubini}. For quantization on the hyperbolic plane, we refer to \cite{ComtetLandauHyperbolic,DUNNE1992233,Grosche:1987de,Grosche:1988um,Balazs:1986uj,Kuperin:1994eh,Kim:2001tw}. The corresponding representation-theoretic perspective originated in the work of the Gel'fand school \cite{GelNai50,GelfandGraevVilenkin1962}.
Quantization using the so-called coherent states dates back to Berezin~\cite{Berezin}; a thorough treatment was given in the seminal book of Perelomov~\cite{PerelomovBook}, cf.~\cite{DeyCoadjoint} for a recent discussion. 
A Berezin-Toeplitz quantization scheme for the hyperbolic case was developed in \cite{KlimekLesniewski1992}.

In this work, we address the following issue. Two-dimensional surfaces mentioned above are known to be Kähler manifolds. 
From the theory of geometric quantization~\cite{woodhouse1992geometric}, we know that Kähler manifolds admit a so-called \textit{holomorphic polarization}. In essence, this means that a phase space which is a Kähler manifold can be quantized into a Hilbert space of square-integrable holomorphic sections of a Hermitian line bundle \cite{Schottenloher2025GeometricQuantization}. In the case of a particle moving on a surface, however, the phase space is not the surface itself, but its cotangent bundle. Moreover, the relation between the Kähler structures on the manifold and on its cotangent bundle is not straightforward \cite{druta2008cotangentbundlesgeneralnatural}. Therefore, for quantization one typically chooses the vertical real polarization, in which the wave functions do not depend on the momenta associated with the cotangent bundle. But we can ask the following question:

\begin{quote}
\textit{Can the Kähler structure of a $2D$ surface itself be used to quantize a free particle moving on that surface in a purely holomorphic manner?}
\end{quote}
The aim of this work is to clarify the answer to this question and to show how this sheds light on the analytic structure of wave functions and the representation-theoretic structure of Hilbert spaces, as well as aids in analyzing the spectral problem for the corresponding quantum Hamiltonians.

The method is based on a geometric idea developed in earlier works \cite{Bykov:2011ai,Bykov:2012am,Bykov:2024tvb}; see \cite{Krivorol2026} for a review. It asserts that if one finds a Lagrangian embedding of the particle's configuration space into coadjoint orbits \cite{Kirillov2001} of its isometry group, and this embedding can be extended to a symplectomorphism with the particle's phase space, then the quantization of the cotangent bundle can be replaced by the quantization of the coadjoint orbits. Furthermore, a Hamiltonian that attains its minimum on the image of the embedding quantizes to an operator that is spectrally equivalent to the Laplace-Beltrami operator on this Lagrangian submanifold.

This observation provides an answer to the question above, as the complex plane, the sphere, and the hyperbolic plane are not only configuration spaces for a free particle but are themselves coadjoint orbits of their respective isometry groups. The relevant Lagrangian embedding has a rather simple structure. Let $\mathbb{M}$ denote one of the spaces: $\mathbb{C}$, $\mathbb{S}^2$, or $\mathbb{H}$. The embedding is given by\footnote{For a discussion of related Lagrangian embeddings see \cite{tyurin2025ample,tyurin2025examples}.}
\begin{align}\label{IntrDiagonalMap}
\mathbb{M}&\hookrightarrow \mathbb{M}\times\mathbb{M}\,,\\
z&\mapsto (z,\smallthickbar{z})\,,\nonumber
\end{align}
where $z$ is a complex coordinate on $\mathbb{M}$. The symplectic form on $\mathbb{M}\times\mathbb{M}$ is a sum of Kirillov-Kostant-Souriau forms with suitable coefficients. Up to some technical details\footnote{In particular, for the sphere, one must consider the limit when the coefficient of the symplectic form on $\mathbb{S}^2\times\mathbb{S}^2$ tends to infinity, as well as quotient out a certain divisor. In case of the hyperbolic plane, the two copies of $\mathbb{H}$ must be equipped with inequivalent $\mathbf{SU}(1,1)$ actions, so we denote them by $\mathbb{H}^+$ and $\mathbb{H}^-$. Consequently, the geometric quantization of $\mathbb{H}^\pm$ yields the positive and negative discrete series representations of $\mathbf{SU}(1,1)$, respectively.}, which we discuss in the text, this leads to the symplectomorphism
\begin{equation}\label{IntrClassIsom}
\mathrm{T}^\ast\mathbb{M} \;``\!\simeq\!"\; \mathbb{M}\times\mathbb{M}\,.
\end{equation}
The quantization of $\mathrm{T}^\ast\mathbb{M}$ yields the $L^2$-space of functions in the variables $z$ and $\smallthickbar{z}$ on $\mathbb{M}$, and the free-particle Hamiltonian in the invariant metric is quantized to the invariant Laplace-Beltrami operator. In contrast, the quantization of $\mathbb{M}$ itself as the Kähler manifold is given by a space of $L^2$-integrable holomorphic sections of a Hermitian line bundle. Consequently, the Hilbert space obtained from quantizing $\mathbb{M}\times\mathbb{M}$ consists of \textit{biholomorphic wave functions} in two variables, which we denote by $z$ and~$w$. In this formalism, the classical Hamiltonian on $\mathbb{M}\times\mathbb{M}$, which replaces under the quantization the Laplace-Beltrami operator on $\mathrm{T}^\ast\mathbb{M}$ and reproduces its eigenvalues, contains a term proportional to $|z-\smallthickbar{w}|^2$ that vanishes on the Lagrangian submanifold we have defined. Moreover, as we will discuss in detail, the standard Laplace operator eigenfunctions from $L^2(\mathbb{M})$ depending on $z$ and $\smallthickbar{z}$ can be obtained from biholomorphic wavefunctions depending on $z$ and $w$ by trivializing the quantum line bundles and restricting to the Lagrangian submanifold $z=\smallthickbar{w}$. 
Thus, our approach elucidates the ``analytical structure'' of eigenfunctions of the Laplace-Beltrami operator on manifolds of constant curvature.

This formalism also provides a transparent geometric and complex-analytic interpretation of certain facts from representation theory. Our analysis reveals its non-trivial nature even for the ``seemingly trivial'' example of the complex plane.
But the most interesting example in our context is the Lobachevsky plane, associated with the group $\mathbf{SL}(2,\mathbb{R})\simeq\mathbf{SU}(1,1)$. For this space, there exists a chain of isomorphisms \cite{Repka,KitaevSL2R}
\begin{equation}\label{IntrThreeIsom}
L^2(\mathbb{H})\simeq \mathcal{D}_p^+\otimes\mathcal{D}_p^-\simeq \int\limits_0^\infty\mathcal{C}_{{1\over 4}+s^2}\,\dif s\,,
\end{equation}
where $\mathcal{D}_p^\pm$ denote the positive and negative discrete series representations with integer parameter $p\geq 2$, and $\mathcal{C}_{q}$ is the principal continuous series representation with Casimir eigenvalue $q\geq {1\over 4}$. In our framework, the first isomorphism can be interpreted geometrically as a quantized version of \eqref{IntrClassIsom} (for the quantization of $\mathbf{SL}(2,\mathbb{R})$ coadjoint orbits, see \cite{Plyushchay1993}). In essence, it arises from realizing $\mathcal{D}_p^\pm$ as spaces of holomorphic functions on the hyperbolic plane, while their tensor product is identified with $L^2(\mathbb{H})$ via a map which is induced by
the `diagonal' map \eqref{IntrDiagonalMap} on Hamiltonian harmonics (but, as we will discuss, an isomorphism between $L^2(\mathbb{H})$ and $\mathcal{D}_p^+\otimes\mathcal{D}_p^-$ defined on general functions is something more involved).

The second isomorphism can be understood through a blend of complex and harmonic analysis. As is well known, the eigenfunctions of the Laplace-Beltrami operator form a basis for the $L^2$-Hilbert space. In the biholomorphic representation, these eigenfunctions are functions of two variables on $\mathcal{D}_p^+\otimes\mathcal{D}_p^-$, but as we show they also depend on an auxiliary parameter $\upgamma$. This parameter $\upgamma$ labels points on the asymptotic boundary at infinity of the hyperbolic plane, where the continuous series representations $\mathcal{C}_{{1\over 4}+s^2}$ are defined. Thus, these eigenfunctions are actually \textit{intertwiners} $\mathcal{D}_p^+\otimes\mathcal{D}_p^-\rightarrow \mathcal{C}_{{1\over 4}+s^2}$ between the ``bulk'' and ``boundary'' representations. They can also be obtained by $\mathbf{SU}(1,1)$ rotations of a reference state and therefore can be viewed as Perelomov coherent states~\cite{PerelomovBook}.

In summary, we believe that our biholomorphic formalism provides a clear conceptual explanation for \eqref{IntrThreeIsom}. Moreover, this formalism also simplifies certain harmonic analysis computations; for instance, it avoids the need to solve difficult partial differential equations when determining the spectrum of the Laplace–Beltrami operator even in the presence of a magnetic field. Furthermore, our approach unifies the analysis for the sphere and the Lobachevsky plane, where all the distinction reduces to differences in the domains of definition of certain ``standard'' wave functions.

\vspace{0.3cm}
The paper is organized as follows. We examine the holomorphic quantization of particles on two-dimensional spaces of constant curvature in a transverse magnetic field. We consider the complex plane $\mathbb{C}$, flat tori $\mathbb{T}^2$, the sphere $\mathbb{S}^2$, and the hyperbolic plane\footnote{The case of hyperbolic surfaces, obtained by quotienting $\mathbb{H}$ by discrete subgroups of $\mathbf{SL}(2,\mathbb{R})$, is technically more involved and is left for future work.} $\mathbb{H}$. These are treated in Sections \ref{PlaneSection}, \ref{SectionTori}, \ref{SphereBigSection} and \ref{H2sec}, respectively. Section~\ref{DiscussionSection} is devoted to a general discussion and an outline of possible future directions.
In Appendix~\ref{zeromagnetic app}, we carefully examine the zero-magnetic-field limit of the wavefunctions on the plane (this is rather subtle as it interpolates between discrete and continuous spectrum). Appendix~\ref{NormAppendix} examines the invariant Hermitian products on the Hilbert spaces in homogeneous coordinates and their translation to inhomogeneous coordinates. Appendix~\ref{AppendixFlatRelation} investigates the relationship between the biholomorphic and standard $L^2$-representations of the Hamiltonian eigenfunctions for free particles. Appendix \ref{scalprodapp} analyzes the structure of scalar products in the hyperbolic case from a symmetry perspective. Finally, Appendix~\ref{Tscalprodapp} details the calculation of the scalar product between eigenfunctions in the hyperbolic case.

\section{The plane  $\mathbb{C}$ and  $\widehat{\mathbf{ISO}}(2)$ orbits}\label{PlaneSection}

Let us start with the discussion of the holomorphic quantization concept in the simplest possible $2D$ geometry, namely the particle on the plane in a transverse magnetic field. Although the quantization of such a system is well known from standard quantum mechanics courses, examining it from a new perspective clarifies the essence of our method and is valuable for constructing generalizations to other geometries.

\subsection{The Landau problem.} 
Consider a free particle on the complex plane~$\mathbb{C}$, with coordinate~$\xi$,   in a transverse magnetic field of strength~$B$. The action for this system is
\begin{equation}
\mathcal{S}[\xi] = \int|\dot{\xi}|^2\dif t +\int\mathcal{A}\qquad\text{where}\qquad\mathcal{A} = \frac{iB\big(\xi\dif\smallthickbar{\xi} - \smallthickbar{\xi}\dif\xi\big)}{2}\,.
\label{FreeAction}
\end{equation}
For simplicity we assume that $B\geq 0$ (if $B<0$, we simply swap $\xi$ and $\smallthickbar{\xi}$\,). Recall that the corresponding quantum Hamiltonian is
\begin{align}\label{LandauHam}
    \widehat{H}_{\mathrm{Landau}}=-\Bigg({\dd\over \dd\xi}-\frac{B \smallthickbar{\xi}\,}{2}\Bigg)\Bigg({\dd\over \dd\smallthickbar{\xi}\,}+\frac{B\xi}{2}\Bigg)\,.
\end{align}

Translational invariance of the original system without magnetic field has an interesting fate when the magnetic field is turned on (cf.~\cite{Davighi:2019ffp}). From the classical perspective, under the shift $\xi \to \xi+a$ the action~(\ref{FreeAction}) is only invariant up to an integral of a total derivative, 
\begin{equation}
\int\mathcal{A}\quad \xrightarrow[\xi\rightarrow\xi+a]{~}\quad \int\big(\mathcal{A}+\dif\zeta\big)\,,\qquad\zeta = \frac{iB\big(a\smallthickbar{\xi}-\smallthickbar{a}\xi\big)}{2}\,,
\end{equation}
whose vanishing depends on the boundary conditions. In turn, the quantum Hamiltonian~(\ref{LandauHam}) will only preserve its form if, in addition to the shift, one performs a gauge transformation
$\widehat{H}_{\mathrm{Landau}}\mapsto e^{-i\zeta}\widehat{H}_{\mathrm{Landau}} e^{i\zeta}$.
This means that the wave function $\Psi\big(\xi, \smallthickbar{\xi}\,\big)$ is multiplied by a phase:
\begin{align}\label{Psitrans1}
    \Psi\big(\xi, \smallthickbar{\xi}\,\big)\; \mapsto\; e^{i\zeta} \Psi\big(\xi+a, \smallthickbar{\xi}+\smallthickbar{a}\big)\,.
\end{align}
Such transformations form a non-Abelian group. Indeed, if one denotes the shift by~$a$ as $g_a$, one finds $g_bg_a=g_ag_b e^{B(\bar{a}b-\bar{b}a)}$. Thus, for $B\neq 0$, in place of the Abelian group of translations one arrives at the three dimensional Heisenberg group. The third generator is the central element, whose action on the wave functions is
\begin{align}\label{Psitrans2}
    \Psi\big(\xi, \smallthickbar{\xi}\,\big)\; \mapsto\; e^{i \theta} \Psi\big(\xi, \smallthickbar{\xi}\,\big)\,.
\end{align}
If one additionally takes into account the rotations in the $\xi$-plane $\mathbb{C}$, one gets the centrally extended $\widehat{\mathbf{ISO}}(2)$ in place of the plane isometry group $\mathbf{ISO}(2)$, where the action of rotations remains intact:
\begin{align}\label{Psitrans3}
    \Psi\big(\xi, \smallthickbar{\xi}\,\big)\; \mapsto\;  \Psi\big(e^{i\phi}\xi, e^{-i\phi}\smallthickbar{\xi}\,\big)\,.
\end{align}

The central extension plays an important role in the foregoing discussion.

\subsection{A `first order' oscillator formulation.} 
To introduce our method, it is beneficial to rewrite the action~(\ref{FreeAction}) in a different form. To this end, consider a seemingly unrelated action
\begin{equation}\label{FlatSpinChainAction}
\mathcal{S}[z,w] = \int\dif t\,\Big(i\lambda\smallthickbar{z}\dot{z} + i(\lambda+B)\smallthickbar{w}\dot{w} - \lambda(\lambda+B)|z-\smallthickbar{w}|^2\Big)\,,
\end{equation}
where $\lambda>0$ is a real  parameter\footnote{As we will see, physically observable quantities such as the Hamiltonian spectrum and wave functions on $\mathbb{C}$ are independent of $\lambda$. For example, we could set $\lambda = 1$ throughout without loss of generality. However, we prefer to keep this parameter for now to illustrate the general logic and connection with the model on the sphere; see section \ref{SphereBigSection}.} of the action, while $z$ and $w$ are complex coordinates on $\mathbb{C}\times\mathbb{C}$. In fact, this action is equivalent to \eqref{FreeAction}. To see this, one makes a linear change of variables of the form
\begin{equation}\label{FlatChangeOfVariables}
\eta = z-\smallthickbar{w}\,,\qquad \xi = \frac{z+\alpha\smallthickbar{w}}{\alpha+1}\qquad\Longrightarrow\qquad z = \xi+\frac{\alpha\eta}{\alpha+1}\,,\qquad \smallthickbar{w} = \xi-\frac{\eta}{\alpha+1}\,.
\end{equation}
The coefficient $\alpha$ is determined by requiring that the term $\smallthickbar{\eta}\dot{\eta}$ be absent from the Lagrangian after the change of variables. This yields
$
\alpha = \left(\frac{\lambda+B}{\lambda}\right)^{1/2}
$. Once the term $\smallthickbar{\eta}\dot{\eta}$ is absent, one can eliminate $\eta$ and $\smallthickbar{\eta}$ via their equations of motion. Up to an integration by parts, the result is precisely the action \eqref{FreeAction}.

Thus, we have represented the free particle in the transverse magnetic field as a system of two ``coupled oscillators''\footnote{In this context, by `oscillators' we mean phase space variables that form the Heisenberg algebra of ladder operators upon quantization, not the potential of a harmonic oscillator.} with phase space $\mathbb{C}\times\mathbb{C}$, equipped with the symplectic form
\begin{equation}
\omega = i \lambda\dif\smallthickbar{z}\wedge \dif z+i(\lambda+B) \dif\smallthickbar{w}\wedge \dif w
\end{equation}
and the Hamiltonian
\begin{align}\label{flatspaceHam}
H_{\mathrm{flat}}=\lambda(\lambda+B)|z-\smallthickbar{w}|^2\,.
\end{align}

The configuration space of the sigma model, i.e.~the $\xi$-plane $\{\eta = 0\}\subset\mathbb{C}\times\mathbb{C}$, coincides with the space of minima $(\mathbb{C})_{\mathrm{min}}$ of the Hamiltonian $H_{\mathrm{flat}}$. Clearly,
\begin{align}\label{CLagrEmbed}
(\mathbb{C})_{\mathrm{min}}\hookrightarrow \mathbb{C}\times\mathbb{C}\,.
\end{align}
An important observation is that for zero magnetic field, $(\mathbb{C})_{\mathrm{min}}$ (given by $z=\smallthickbar{w}$) is a Lagrangian submanifold. However, for $B\neq 0$, the image of $(\mathbb{C})_{\mathrm{min}}$ is symplectic, and the restriction of the symplectic form is given by
\begin{align}\label{SymplecticB}
\Big(i \lambda\,\dif\smallthickbar{z}\wedge \dif z+i(\lambda+B) \dif\smallthickbar{w}\wedge \dif w\Big)\Big|_{z=\smallthickbar{w}}=iB\, \dif\smallthickbar{w}\wedge \dif w\,.
\end{align}

This geometric picture represents the simplest particular case of a correspondence between 1$D$ sigma models (free particles on manifolds) and `classical spin chains' (models whose phase spaces are products of coadjoint orbits) \cite{Bykov:2012am,Bykov:2024tvb}, see also \cite{Krivorol2026}. 

\subsection{$\widehat{\mathbf{ISO}}(2)$ coadjoint orbits.}\label{SectionISOcoadjointOrbits} 
In~(\ref{CLagrEmbed}) the r.h.s.~may be viewed as the product of two coadjoint orbits of the Heisenberg group; see section 4.3 in \cite{Lahlali:2024qnk}. Moreover, these orbits can also be regarded as coadjoint orbits of a ``true'' symmetry group of $(\mathbb{C})_{\mathrm{min}}$, namely the centrally extended $\mathbf{ISO}(2)$. For convenience and completeness of the exposition, in this section we present the derivation of this fact.

Recall that the $\widehat{\mathfrak{iso}}(2)$ algebra has the nontrivial commutation relations
\begin{equation}
[A,A^\dagger] = B\,,\qquad [A,H] = A\,,\qquad [A^\dagger,H] = -A^\dagger\,, 
\end{equation}
where we denote the $\mathfrak{iso}(2)$ abstract generators as $A$, $A^\dagger$ and $H$, and $B$ as the central element. This is the form of the algebra as it appears in our work. But in this section it is more transparent to work in a different real basis
\begin{equation}
[J,X] = Y\,,\qquad [J,Y] = -X\,,\qquad [Y,X] = \frac{1}{2}C\,.
\end{equation}
This means that the Kirillov-Kostant-Souriau Poisson bracket on $\widehat{\mathfrak{iso}}(2)^\ast$ has the structure
\begin{equation}
\{j,x\} = y\,,\qquad \{j,y\} = -x\,,\qquad \{y,x\} = \frac{1}{2}c\,,
\end{equation}
where $x$ and $y$, $j$ and $c$ are coordinates on $\widehat{\mathfrak{iso}}(2)^\ast$ viewed as $\mathbb{R}^4$.

There is a well-known fact that the symplectic leaves of the Kirillov-Kostant-Souriau Poisson bracket on the dual space of the Lie algebra are precisely the coadjoint orbits. Let us prove that there exist  symplectic leaves in $\widehat{\mathfrak{iso}}(2)^\ast$ symplectomorphic to the complex planes with the standard symplectic form. The one obvious Casimir function is the ``central'' coordinate $c$ and we fix it to a non-zero constant. The second Casimir function is less obvious and has the form 
$R^2 := x^2+y^2-cj$
which we also fix to be constant. This defines a paraboloid, which can be projected onto the $(x,y)$-plane with the induced standard symplectic structure proportional to $c^{-1}$.

Thus, the centrally extended $\mathbf{ISO}(2)$ group shares the coadjoint orbits with the three dimensional Heisenberg group. It can be explained by the fact that $\widehat{\mathfrak{iso}}(2)$ contains the Heisenberg algebra as subalgebra spanned by $A$, $A^\dagger$ and $B$, whereas the extra generator $H$ can be constructed\footnote{Equivalently, the generator $J$ may be chosen to be proportional to $X^2 + Y^2$.} as the operator $A^\dagger A$ in the universal enveloping algebra. 

Another interesting case is when the central extension vanishes, i.e.~when we set $c=0$, so that one deals with coadjoint orbits of $\textbf{ISO}(2)$. In this case the symplectic leaves are given by $R^2=x^2+y^2=\mathrm{const}$, so that they are cylinders parametrized by $(x, y, j)$. Each such cylinder should be thought of as the cotangent bundle $\mathrm{T}^\ast \mathbb{S}^1$, so that upon quantization one gets spaces of functions on the circle, as we shall see later.

\subsection{A quick aside: Fock space from orbit quantization.} 
Let us clarify in this section how to quantize the model mechanical system whose Lagrangian is $\mathcal{L} =  i\smallthickbar{z}\dot{z}$, where $z = x+iy$ is the complex coordinate on the complex plane $\mathbb{C}$. From the geometric point of view this system is the mechanics on the $\widehat{\mathbf{ISO}}(2)$ coadjoint orbit defined by the Kirillov-Kostant-Souriau symplectic structure, see the discussion above. 

We will assume that the elements of our Hilbert space $\mathscr{B}(\mathbb{C})$ are the holomorphic functions $f(z)$ on $\mathbb{C}$. Due to the symplectic structure of the Lagrangian, the coordinate $z$ should quantize into the coordinate operator, whereas $p_z:=i\smallthickbar{z}$ should quantize into the momentum operator. For example, we can work in the coordinate or momentum representations, whose respective quantization maps from functions to operators are
\begin{align}\label{FlatQuantizRules}
    z\mapsto z\,,\quad \smallthickbar{z}\mapsto -{\dd\over \dd z}\qquad\text{or}\qquad z \mapsto \frac{\partial}{\partial z}\,,\quad \smallthickbar{z}\mapsto z\,.
\end{align}
In this work we prefer the second choice.

Next we need to define the Hilbert space structure, i.e.~to define the scalar product: 
\begin{align}
    (h, f)=\int\,\dif^2z\,\smallthickbar{h}(\smallthickbar{z}) f(z)\,\rho(z,\smallthickbar{z})\,,
\end{align}
where $\rho(z,\smallthickbar{z})$ is the measure. 
Our main requirement is that $\widehat{\mathbf{ISO}}(2)$ be represented by \textit{unitary} operators in this Hilbert space. In other words, for every $g\in \widehat{\mathbf{ISO}}(2)$ one should have $(g\circ h, g\circ f)=(h, f)$. Assuming an action of the form $g\circ f(z)=\omega_g(z)\,f(g\circ z)$ for some function $\omega_g(z)$, one finds the following unique answer: $\rho(z, \smallthickbar{z})=e^{-|z|^2}$ and
\begin{align}
    &g=e^{i\theta H}\quad\rightarrow\quad g\circ f(z)=f\big(e^{i\theta}z\big)\,,\\
    &g=e^{i \alpha C}\quad\rightarrow\quad g\circ f(z)=e^{i\alpha} f(z)\,,\\
    &g=e^{i(u A^\dagger+\bar{u}A)}\quad\rightarrow\quad g\circ f(z)=e^{-{1\over 2}|u|^2}e^{-\bar{u}z}f(z+u)\,.
\end{align}
Thus the inner product is given by the \textit{Bargmann-Fock formula}
\begin{align}\label{BFMeasure}
    (h, f)_{\mathscr{B}(\mathbb{C})}=\int\,\dif^2z\,\smallthickbar{h}(\smallthickbar{z}) f(z)\,e^{-|z|^2}\,.
\end{align}
Note that the invariant measure includes the factor $e^{-\mathscr{K}}$, where $\mathscr{K} = |z|^2$ is the Kähler potential for the complex plane, see \cite{DUNNE1992233,PerelomovBook}. Some of the isometries of the plane shift the potential by the (holomorphic) K\"ahler transformations, and these in turn arise as the twist factors $\omega_g(z)$ in the above formulas.

The Hilbert space $\mathscr{B}(\mathbb{C})$ is isomorphic to the Fock space with its standard inner product, the map between the two being given by
$
    f(z) \rightsquigarrow  f(a^\dagger)|0\rangle\,. 
$

\subsection{Holomorphic quantization of the particle.} \label{SubsectionFlatHolomorphicQuant}
Let us now quantize the theory~\eqref{FlatSpinChainAction}. Analogously with \eqref{FlatQuantizRules}, we choose the following representation
\begin{equation}\label{FlatQuantizationRules}
z\mapsto \lambda^{-1}\frac{\partial}{\partial z}\,,\quad \smallthickbar{z}\mapsto z\,,\qquad w\mapsto(\lambda+B)^{-1}\frac{\partial}{\partial w}\,,\quad \smallthickbar{w}\mapsto w\,.
\end{equation}
The quantum Hamiltonian is\footnote{Here the possible ordering ambiguity amounts to adding a constant to the Hamiltonian.} 
\begin{equation}\label{QuantumFlatHamiltonian}
\widehat{H} = \lambda(\lambda+B)\bigg(z -  (\lambda+B)^{-1}\frac{\partial}{\partial w}\bigg)\bigg(\lambda^{-1}\frac{\partial}{\partial z}-w\bigg):=A^\dagger A\,,
\end{equation} 
where we introduced generalized ladder operators
\begin{equation}\label{AAdaggerDef}
A^\dagger =\sqrt{\lambda(\lambda+B)} \bigg(z -  (\lambda+B)^{-1}\frac{\partial}{\partial w}\bigg)\,,\qquad A = \sqrt{\lambda(\lambda+B)} \bigg(\lambda^{-1}\frac{\partial}{\partial z}-w\bigg)
\end{equation}
which commute as
\begin{equation}\label{AAdaggerCommutation}
[A,A^\dagger] = B\,.
\end{equation}
Clearly, $A, A^\dagger, \widehat{H}$ and the central element $B$ also constitute the Lie algebra of $\widehat{\mathfrak{iso}}_2$. 

The spectral problem is formulated as
\begin{align}\label{planemagneticeqn}
    \lambda(\lambda+B)\left(z-(\lambda+B)^{-1}{\dd \over \dd w}\right) \left(\lambda^{-1}{\dd\over \dd z}- w\right)\Phi(z, w)=E \Phi(z, w)\,.
\end{align}
Notice that this is the differential equation for an entire function in $\mathbb{C}\times \mathbb{C}$. The space of such functions equipped with the measure 
\begin{equation}
\big(h(z,w),f(z,w)\big)_\mathscr{B} = \int\dif^2 z\,\dif^2 w\, \overline{h(z,w)}f(z,w) \,e^{-\lambda|z|^2-(\lambda+B)|w|^2}
\end{equation}
is \textit{the Bargmann–Fock space} which we denote\footnote{In this notation the space with the measure~(\ref{BFMeasure}) is $\mathscr{B}^1(\mathbb{C})$.} as $\mathscr{B}^{\lambda}(\mathbb{C})\otimes\mathscr{B}^{\lambda+B}(\mathbb{C})$. It follows from section \ref{SectionISOcoadjointOrbits} that $\lambda^{-1}$ and $(\lambda+B)^{-1}$ can be viewed as central charges of the corresponding Heisenberg representation. As will become apparent, the tensor product spaces $\mathscr{B}^{\lambda}(\mathbb{C})\otimes\mathscr{B}^{\lambda+B}(\mathbb{C})$ exhibit fundamentally different structures depending on whether $B=0$ or $B\neq 0$.

\subsubsection{Dual $\widehat{\mathfrak{iso}}_2$ algebra.}  
In addition, the model is clearly invariant with respect to the diagonal $\widehat{\mathfrak{iso}}_2$ action. The generators of this Lie algebra are $P$, $P^\dagger$, and $P^\dagger P$, where $P$ is the momentum corresponding to shifts of $\xi$. Explicitly,
\begin{align}
P=(\lambda+B)w - \frac{\partial}{\partial z}\,,\quad\quad P^\dagger=\frac{\partial}{\partial w} - \lambda z\,.
\end{align}
Indeed, since according to \eqref{FlatChangeOfVariables},
we have $[P^\dagger, \xi]=1$
and similarly $[-P, \xi^\dagger]=1$. Notice that the commutator of the momenta is now non-vanishing:
\begin{align}\label{PPdagalg}
[P, P^\dagger]=-B\,.
\end{align}
An important observation is that $[P, A]=[P, A^\dagger]=0$, meaning the two copies of $\widehat{\mathfrak{iso}}_2$ centralize each other. The presence of two mutually centralizing copies of the symmetry algebra is a general feature of the models studied in this paper. This phenomenon corresponds to what is known as a \textit{Howe dual pair} \cite{Howe1979seriesAI}; see also \cite{Basile:2020gqi} for further examples based on oscillators.

\subsection{Case of vanishing magnetic field.} 
First consider the case of vanishing magnetic field, $B=0$. In this case the formal solutions of \eqref{planemagneticeqn} have the form 
\begin{align}
    \Phi_{\alpha, \beta}(z,w)=\exp\big({\lambda wz+\alpha z+\beta w }\big)\,,\quad\quad E=-\alpha\beta\,.
\end{align}
So far $\alpha$ and $\beta$ are arbitrary complex numbers and the eigenvalue $E$ is formally complex. However,  we should also require that the eigenfunctions be $\delta$-normalizable. Let us compute the scalar product $(\Phi_{\alpha_2, \beta_2}, \Phi_{\alpha_1, \beta_1})_{\mathscr{B}}:=(\Phi_{2}, \Phi_{1})$ of two such functions:
\begin{align}
\nonumber    (\Phi_{2}, \Phi_{1})&=\int\dif^2z\,\dif^2w\,\exp\Big({-\lambda |z|^2-\lambda |w|^2+\lambda\, wz+\lambda \,\smallthickbar{w}\smallthickbar{z}+\alpha_1 z+\beta_1 w+\smallthickbar{\alpha}_2 \smallthickbar{z}+\smallthickbar{\beta}_2 \smallthickbar{w}}\Big)=\\&=\int\dif^2z\,\dif^2w\,\exp\Big({-\lambda |z-\smallthickbar{w}|^2+\alpha_1 z+\beta_1 w+\smallthickbar{\alpha}_2 \smallthickbar{z}+\smallthickbar{\beta}_2 \smallthickbar{w}}\Big)=\\
&=\{z\to z+\smallthickbar{w}\}={\pi\over \lambda}\,e^{{1\over \lambda}\alpha_1\smallthickbar{\alpha}_2} \,\int\,\dif^2w\,\exp\Big({(\beta_1+\smallthickbar{\alpha}_2)w+(\alpha_1+\smallthickbar{\beta}_2)\smallthickbar{w}}\Big)\,.\nonumber
\end{align}
The latter integral is well-defined and gives a 2$D$ $\delta$-function only if $\beta_i=-\smallthickbar{\alpha}_i, \;i=1, 2$. In this case our `harmonics' may be relabeled as $\Phi_\alpha$, and they take the form 
\begin{align}\label{FlatPlaneWaves}
    \Phi_\alpha=e^{-{1\over 2\lambda}|\alpha|^2}\exp\big({\lambda wz+\alpha z-\smallthickbar{\alpha}w}\big)\,,\quad\quad E=|\alpha|^2\,,
\end{align}
where we have inserted an extra  factor to get the canonical normalization of the scalar product:
$
    (\Phi_{\alpha_2}, \Phi_{\alpha_1})_\mathscr{B}\sim \delta^2(\alpha_1-\alpha_2)\,.
$  
We are now in a position to elaborate the isomorphism\footnote{For a related discussion of this isomorphism and its connection to Weyl quantization, see~\cite[section 1.2]{Ors}.} 
\begin{align}\label{BBL2iso}
    \mathscr{B}^{\lambda}(\mathbb{C})\otimes\mathscr{B}^{\lambda}(\mathbb{C})\simeq L^2(\mathbb{C})
\end{align}
of the two Hilbert spaces.  Suppose we are given a function $\Phi(z,w)\in \mathscr{B}^{\lambda}(\mathbb{C})\otimes\mathscr{B}^{\lambda}(\mathbb{C})$. By the above results, it may be decomposed in a basis of Hamiltonian eigenfunctions~\eqref{FlatPlaneWaves} as
\begin{align}\label{AnalogOfFourier}
    \Phi(z,w)=\int\,\dif^2\alpha\,\Psi(\alpha, \smallthickbar{\alpha})\,\Phi_{\alpha}(z,w)\,,\quad \textrm{where}\quad \Psi(\alpha, \smallthickbar{\alpha})\in L^2(\mathbb{C})\,.
\end{align}
Then the map $\Phi\mapsto \Psi(\alpha, \smallthickbar{\alpha})$ clearly is an isometry of the two Hilbert spaces, i.e.~there is the ``Parseval's identity'' of the form
\begin{equation}
\big\lVert\Phi(z,w)\big\rVert^2_{\mathscr{B}^{\lambda}(\mathbb{C})\otimes\mathscr{B}^{\lambda}(\mathbb{C})} = \big\lVert \Psi(\alpha,\smallthickbar{\alpha})\big\rVert_{L^2(\mathbb{C})}^2\,.
\end{equation}
This isomorphism is important because it connects the Hilbert spaces of the particle on the plane quantized in different coordinates (or, mathematically, in two different polarizations).

Let us also emphasize the representation-theoretic content of the isomorphism~(\ref{BBL2iso}), which should be viewed as an isomorphism of $\widehat{\mathbf{ISO}}(2)$ representations, meaning that it is equivariant w.r.t.~the group action. To this end, note that the space of `plane waves' of the form~(\ref{FlatPlaneWaves}) with fixed energy is isomorphic to the space of functions on a circle $|\alpha|^2=E$, since for any such function $f$ one can form a linear combination
\begin{align}\label{Phithetaintegral}
    \Phi_E(z,w)=
    \int\,\dif^2\alpha \,\delta\big(|\alpha|^2-E\big)\,f(\alpha)\,\exp\big({\lambda wz+\alpha z-\smallthickbar{\alpha} w}\big)\,.
\end{align}
The space of such functions, equipped with the standard inner product, constitutes an irreducible unitary representation $\mathcal{R}^{\mathbf{ISO}(2)}_{E}$ of $\mathbf{ISO}(2)$~\cite[Chapter 4]{VilenkinKlimyk1}. One may also view these functions as arising from the quantization of the coadjoint orbits of $\mathbf{ISO}(2)$ that are cylinders $\mathrm{T}^\ast \mathbb{S}^1$, as mentioned at the end of section~\ref{SectionISOcoadjointOrbits}. Thus, we may refine the isomorphism~(\ref{BBL2iso}) as follows:
\begin{align}\label{BBL2isoRep}
\mathscr{B}^{\lambda}(\mathbb{C})\otimes\mathscr{B}^{\lambda}(\mathbb{C})\simeq L^2(\mathbb{C})\simeq \int\limits_0^\infty\,\mathcal{R}^{\mathbf{ISO}(2)}_{E}\dif E\,.
\end{align}
Here the l.h.s.~is the tensor product of $\widehat{\mathbf{ISO}}(2)$ irreducible representations with central charge $\lambda^{-1}$, whereas the r.h.s.~is a sum of irreducible representations with vanishing central charge.

\subsection{Magnetic wavefunctions.} Here we will find the  eigenfunctions in a non-zero magnetic field. We start from the ground state solution $\Phi_0$ corresponding to $E=0$ and satisfying $\left(\lambda^{-1}\dd_z- w\right)\Phi_0(z,w)=0$. Clearly, the  solution is
\begin{align}\label{Phi0}
    \Phi_0(z,w)=f(w)\,\exp\big(\lambda zw\big)\,.
\end{align}
Since there is a copy of $\widehat{\mathfrak{iso}}_2$ algebra \big(generated by $P, P^\dagger$, cf.~(\ref{PPdagalg})\big), which commutes with the Hamiltonian, the space of eigenstates of any (fixed) energy furnishes a representation of this algebra. It is thus natural to look for a ground state with the property $P^\dagger \Phi_0=\beta \Phi_0$, with $\beta \in \mathbb{C}$. This gives
\begin{align}
    \Phi_0(z,w|\beta)=\exp\big(\lambda zw+\beta w\big)
\end{align}
One can act on this state with the operator $P$ any number of times and get other states with the same (zero) energy, which together span a Fock module that we call $\mathscr{B}_0^{B}(\mathbb{C})$.

To get an excited state one may now act on this ground state $\Phi_0(z,w|\beta)$ with the raising operator $A^\dagger = z-(\lambda+B)^{-1}\partial_w$  sufficiently many times. For a state at level $n$ one gets
\begin{align}\label{PhiNstate}
    \Phi_n(z,w|\beta)=(zB-\beta)^{n}\,\exp\big(\lambda zw+\beta w\big)\,.
\end{align}
Again, one can act on this state with $P, P^2, \ldots$ generating a Fock module that we label\footnote{The representations for different $n$ are all isomorphic, so the subscript is just used for book-keeping purposes.} $\mathscr{B}_n^{B}(\mathbb{C})$. States belonging to this multiplet have energy 
\begin{align}\label{Landauenergy}
  E=Bn\,,  \quad\quad n=0, 1, \,\ldots\,, \infty\,,
\end{align}
which simply follows from the commutation relation~\eqref{AAdaggerCommutation}.

Here the analogue of the decomposition~(\ref{BBL2isoRep}) of the product of irreducible representations of $\widehat{\mathbf{ISO}}(2)$ is
\begin{align}
    \mathscr{B}^{\lambda}(\mathbb{C})\otimes\mathscr{B}^{\lambda+B}(\mathbb{C})\simeq \bigoplus_{n=0}^\infty\; \mathscr{B}_n^{B}(\mathbb{C})\simeq L^2_{(B)}(\mathbb{C})\,.
\end{align}
Here 
 $L^2_{(B)}(\mathbb{C})$ is the space of $L^2$-functions $f$ on $\mathbb{C}$ with the standard scalar product and an action of $\widehat{\mathbf{ISO}}(2)$ given in~(\ref{Psitrans1})-(\ref{Psitrans3}) (notice that it preserves the scalar product).

In Appendix~\ref{zeromagnetic app} we discuss how the eigenfunctions~(\ref{PhiNstate}) converge to the plane waves~(\ref{FlatPlaneWaves}) in the limit $B\to 0$.

\section{The torus $\mathbb{T}^2$ and quotient of $\widehat{\mathbf{ISO}}(2)$ orbits}\label{SectionTori}

As an application of our formalism let us solve the quantum free particle on a flat torus. In other words, we calculate the spectrum of the Laplace-Beltrami operator on tori equipped with a flat Riemannian metric. We assume that the torus is defined by a lattice generated by vectors $1$ and $\tau$ in $\mathbb{C}$, i.e.~$\mathbb{T}^2\simeq\frac{\mathbb{C}}{\mathbb{Z}+\tau\mathbb{Z}}$. We denote the corresponding lattice as $\mathsf{\Lambda}$. Note that the phase space $\mathrm{T}^\ast\mathbb{T}^2$ of the problem is realized as $\frac{\mathbb{C}\times\mathbb{C}}{\mathbb{Z}+\tau\mathbb{Z}}$ with a natural symplectic form inherited from $\mathbb{C}\times\mathbb{C}$, where the lattice $\mathsf{\Lambda}$ acts on $(z,w)\in\mathbb{C}\times\mathbb{C}$ diagonally:
\begin{equation}
z\rightarrow z+1\,,\quad w\rightarrow w+1\qquad\text{and}\qquad z\rightarrow z+\tau\,,\quad w\rightarrow w+\smallthickbar{\tau}\,.
\end{equation}
Moreover, in the absence of magnetic field one has the Lagrangian embedding
\begin{align}
    \mathbb{T}^2\hookrightarrow \frac{\mathbb{C}\times\mathbb{C}}{\mathbb{Z}+\tau\mathbb{Z}}\,,
\end{align}
which is the analogue of~(\ref{CLagrEmbed}). For the sake of simplicity let us also fix $\lambda = 1$ in this section.

\subsection{Zero magnetic field.}
We start with the case $B = 0$ and define the shift operators as
\begin{equation}
\mathsf{T}_v = e^{v P^\dagger-\widebar{v} P}\,,\quad\quad \textrm{where}\quad\quad P = w - \frac{\partial}{\partial z}\,,\qquad P^\dagger =\frac{\partial}{\partial w} - z\,.
\end{equation}
They are quantum analogues of the transformations
\begin{equation}
\xi\rightarrow\xi+v\,,\qquad\smallthickbar{\xi}\rightarrow\smallthickbar{\xi}+\smallthickbar{v}\,.
\end{equation}
Since $[P,P^\dagger] =[P,A] = [P,A^\dagger] = 0$ in the absence of magnetic field, the shift operators commute with each other as well as with the Hamiltonian $\widehat{H} = A^\dagger A$. 

On the torus, we search for the joint eigenfunctions of the  operators $\mathsf{T}_1$ and $\mathsf{T}_\tau$. In other words, we ask for the functions \eqref{FlatPlaneWaves}
such that\footnote{One could as well consider `Bloch waves', i.e. states satisfying $\mathsf{T}_1|\Phi\rangle=e^{i\alpha_1}|\Phi\rangle$, $\mathsf{T}_\tau|\Phi\rangle=e^{i\alpha_2}|\Phi\rangle$, where $\alpha_{1,2}$ are the so-called quasi-momenta \cite{Dereli:2021wqj}. For the sake of simplicity we restrict here to the case $\alpha_{1,2} = 0$.} 
\begin{equation}\label{TorusConstraints}
\mathsf{T}_1 \Phi_\beta(z,w) = \Phi_\beta(z,w)\,,\qquad\text{and}\qquad \mathsf{T}_\tau\Phi_\beta(z,w) = \Phi_\beta(z,w)\,.
\end{equation}
This imposes restrictions on the allowed values of $\beta$. Concretely, the allowed values should satisfy 
\begin{equation}
\beta - \smallthickbar{\beta} = 2\pi i n\,,\qquad
\tau\beta - \smallthickbar{\tau}\smallthickbar{\beta} = 2\pi i m\,,\qquad n,m\in\mathbb{Z}\,.
\end{equation}
Thus, the possible eigenvalues $E=|\beta|^2$ of the Hamiltonian are
\begin{equation}\label{TorusSpectrum1}
E = 4\pi^2\Bigg(n^2+\frac{(m-\tau_R n)^2}{\tau^2_I}\Bigg)\,,\qquad\tau:=\tau_R+i\tau_I\,.
\end{equation}
This formula can be rewritten in a more familiar form in terms of the dual lattice $\mathsf{\Lambda}^\star$. It can be generated by dual basis vectors
\begin{equation}
v_1^\ast = 1-i\bigg(\frac{\tau_R}{\tau_I}\bigg)\,,\qquad v_2^\ast = \frac{i}{\tau_I}\,.
\end{equation}
Every dual vector can be represented as $\vec{k} = n v_1^\ast + m v_2^\ast$. By using this notation, the formula \eqref{TorusSpectrum1} can be rewritten more concisely as
\begin{equation}
E = 4\pi^2 |\vec{k}|^2\,.
\end{equation}
This reproduces the well-known result for the spectrum of the  Laplace-Beltrami operator on a torus~\cite{Milnor}.

\subsection{Non-zero magnetic field.}
Next we pass over to the case $B\neq 0$, which was studied in~\cite{Novikov1, novikov1981magnetic}, see also \cite{Dereli:2021wqj,OnofriTorus,Fubini}.
The magnetic shift operators by periods of the torus are
\begin{align}
    \mathsf{T}_1:=e^{P^\dagger-P}\,,\quad\quad \mathsf{T}_{\tau}:=e^{\tau P^\dagger-\bar{\tau} P}\,,
\end{align}
where the symbols of $P$ and $P^\dagger$ are modified as
\begin{equation}
P = (1+B)w - \frac{\partial}{\partial z}\,,\qquad P^\dagger =\frac{\partial}{\partial w} - z\,.
\end{equation}
One easily shows that they obey the \textit{magnetic group relation} \cite{Zak:1964zz} 
\begin{align}
\mathsf{T}_1 \mathsf{T}_\tau=\mathsf{T}_\tau \mathsf{T}_1 \,e^{-B(\tau-\bar{\tau})}\,.   
\end{align}
The requirement that $\mathsf{T}_1$ and $\mathsf{T}_{\tau}$ commute leads to the Dirac quantization condition of magnetic flux: 
\begin{align}\label{QuntizationOfB}
     2B\,\mathfrak{I}(\tau) =2\pi i  n\,,\quad\quad n\in \mathbb{Z}\,,
\end{align}
where $\mathfrak{I}(\bullet)$ means the imaginary part.
In other words, the Dirac quantization condition may be formulated as
\bea
iB\int\limits_{\mathbb{T}^2} \dif\xi\wedge \dif\smallthickbar{\xi}=2\pi n\,,\quad\quad n\in \mathbb{Z}\,.
\eea

\subsubsection{Lowest Landau level.}
Let us discuss the concrete form of the torus wavefunctions in the case of the zero energy in our biholomorphic form.
The first torus constraint \eqref{TorusConstraints} is given by 
\begin{equation}
\mathsf{T}_1\Phi_0(z,w) = \Phi_0(z,w) = f(w)e^{zw}\quad\Longleftrightarrow\quad f(w+1) = \exp\bigg(\frac{B}{2}+Bw\bigg)f(w)
\end{equation}
which can be solved by
\begin{equation}\label{MagneticWaves2}
\Phi_0(z,w) = \exp\bigg({\frac{B}{2}w^2 + zw}\bigg)\phi(w)\,,
\end{equation}
where $\phi(w)$ is a periodic function, i.e.~$\phi(w+1) = \phi(w)$.

Let us also analyze the second constraint in \eqref{TorusConstraints}, namely $\mathsf{T}_\tau\Phi_0(z,w) = \Phi_0(z,w)$. 
The action of $\mathsf{T}_\tau$ on \eqref{MagneticWaves2} is given by
\begin{equation}
\phi(w+\tau) = \exp\Big(-iB\mathfrak{I}(\tau)\big(2w+\tau\big)\Big)\phi(w) = \exp\Big(-i\pi n\big(2w+\tau\big)\Big)\phi(w)\,,
\end{equation}
where we use the quantization condition \eqref{QuntizationOfB}. The solutions of this equation are the Jacobi theta functions of the level $n$ \cite{FlorentinoMouraoNunes2006,Dereli:2021wqj},
\begin{equation}
\vartheta_{n,k}(w|\tau) = \sum_{m\in\mathbb{Z}}\exp\Bigg(i\pi n\tau\bigg(m+\frac{k}{n}\bigg)^2+2\pi i n w \bigg(m+\frac{k}{n}\bigg)\Bigg)\,,\quad k=0,1,\ldots,n-1\,.
\end{equation}
They are the global sections $H^0\big(\mathbb{T}^2,\mathfrak{L}^n\big)$ of a certain magnetic line bundle $\mathfrak{L}^n$ over the torus, and the dimension of the space of such global sections space is precisely $n$. This recovers the standard result that the degeneracy of the lowest Landau level on the torus is proportional to the magnetic field $B$. To summarize, the basis in the space of ground states with period $\tau$ and magnetic flux \eqref{QuntizationOfB} is furnished by the functions  
\begin{equation}
\Phi^{n,k}_{0}(z,w) = \exp\bigg(\frac{B}{2}w^2+zw\bigg)\vartheta_{n,k}(w|\tau)\,.
\end{equation}
The excited states can be obtained by the action of $A^\dagger$ defined by \eqref{AAdaggerDef} on $\Phi^{n,k}_{0}(z,w)$ which leads to an additional $z$-dependent prefactor. Such excited states are also eigenfunctions of $\mathsf{T}_1$ and $\mathsf{T}_\tau$ since $A^\dagger$ commutes with $P$ and $P^\dagger$. 

Note that the eigenfunctions can also be expressed through the $\sigma$-function rather than the $\vartheta$-functions, see \cite{novikov1981magnetic}.

\section{The sphere $\mathbb{S}^2$ and $\mathbf{SU}(2)$ orbits}\label{SphereBigSection}
In the previous section, we discussed the quantization of a particle in a transverse magnetic field on flat spaces, such as planes and flat tori. Now, we apply the holomorphic method of quantization to the space of positive curvature. 
In other words, our goal here is to quantize the particle on the sphere $\mathbb{S}^2$ in a way similar to~Section~\ref{PlaneSection}. Concretely, we will reproduce the spectrum of magnetic harmonics on a sphere from elementary reasons of complex analyticity of the wavefunctions, cf.~also~\cite{Bykov:2024tvb} and~\cite{Krivorol2026}. 

\subsection{Classical model.}
Let us consider the classical model
\begin{equation}\label{ActionSphere}
\mathcal{S}[z,w] = \int\dif t\,\Big(i\smallthickbar{z}\circ\dot{z}+i\smallthickbar{w}\circ\dot{w} - \big(\smallthickbar{z}\circ w\big)\big(\smallthickbar{w}\circ z\big)\Big)\,,\qquad |z|^2 = p\,,\quad |w|^2 = p+\mathfrak{q}\,,
\end{equation}
where $z$ and $w$ are complex vectors in $\mathbb{C}^2$ of fixed lengths and $\circ$  denotes the standard sesquilinear product in $\mathbb{C}^2$, for example $\smallthickbar{z}\circ w:=\smallthickbar{z}_{\,\alpha} w^\alpha = \smallthickbar{z}_{\,1} w^1+\smallthickbar{z}_{\,2} w^2$.
Evidently, the phase space of such mechanical system is $\mathbb{S}^2\times \mathbb{S}^2$. 

If one writes the above action using the inhomogeneous coordinates\footnote{Here and hereafter, with a slight abuse of notation, we use the same letters to denote both homogeneous and inhomogeneous coordinates, their meaning being clear from the context.} on the spheres $z=[z^1:z^2]\mapsto [z,1]$ and $w=[w^1:w^2]\mapsto [1,-w]$, the Hamiltonian takes the form similar to~(\ref{flatspaceHam}):
\begin{align}\label{S2Ham}
    H_{\mathrm{sphere}}=p(p+\mathfrak{q})\frac{|z-\smallthickbar{w}|^2}{(1+|z|^2)(1+|w|^2)}\,,\qquad z,w\in\mathbb{C}\,.
\end{align}
The minimum of the Hamiltonian, corresponding to $z=\smallthickbar{w}$, or $\smallthickbar{z}\circ w=0$ in the homogeneous coordinates as in (\ref{ActionSphere}), provides an embedding
\begin{align}
    \big(\mathbb{S}^2\big)_{\mathrm{min}}\hookrightarrow \mathbb{S}^2\times \mathbb{S}^2\,,
\end{align}
which is Lagrangian in the absence of magnetic field (and symplectic otherwise) with respect to the symplectic form
\begin{equation}\label{SymplecticFormSphere}
\omega = i \dif\smallthickbar{z}_{\,\alpha}\wedge \dif z^\alpha+i \dif\smallthickbar{w}_\alpha\wedge \dif w^\alpha
\end{equation}
on the product of the spheres. It can be checked, in full accordance with \eqref{SymplecticB}, that in the case of the monopole  field the pullback of the symplectic form to the Lagrangian submanifold is proportional to $\mathfrak{q}$.

As shown in~\cite{Bykov:2024tvb}, in the limit $p\rightarrow\infty$ this action converges to the $1D$ sigma model with the sphere target space constructed from the metric\footnote{A similar calculation is carried out below in Section~\ref{H2sec} for the case of the hyperbolic plane~$\mathbb{H}$. Curiously, in that case no limit is necessary.}
\begin{equation}
\dif s^2 = \frac{\dif\xi\dif\smallthickbar{\xi}}{\big(1+|\xi|^2\big)^2}\,,\qquad \xi\in \mathbb{C}\,.
\end{equation}
One can also prove that for $\mathfrak{q}=0$ there 
is a symplectomorphism of the form
\begin{equation}\label{SymplWithSphere}
\mathrm{T}^\ast\big(\mathbb{S}^2\big)_{\mathrm{min}}\simeq\lim_{p\rightarrow\infty}\Bigg(\frac{\mathbb{S}^2\times\mathbb{S}^2}{\big\{\varepsilon_{\alpha\beta}z^\alpha w^\beta = 0\big\}}\Bigg)\,,
\end{equation}
see \cite{Bykov:2024tvb,Krivorol2026}.
Thus, the action \eqref{ActionSphere} is the ``positive constant curvature'' analog of the action \eqref{FlatSpinChainAction}. In this way, the particle’s phase space on the sphere is represented as a product of two $\mathbf{SU}(n)$ coadjoint orbits (which are also spheres) in the limit where their radii become large.

\subsection{Holomorphic quantization.}\label{SphereHolomorphicQuant} The standard calculation of the particle spectrum on the sphere $\mathbb{S}^2$ involves solving a second-order elliptic partial differential equation. This originates from the conventional quantization of the cotangent bundle over the sphere, which relies on choosing a real (vertical) polarization (or, in physical terms, the coordinate representation). The idea that simplifies the quantum problem is to employ holomorphic quantization. This is possible because, as discussed above, the phase space of the particle on the sphere is ``almost symplectomorphic'' to a product of two spheres, each of which is a coadjoint orbit of $\mathbf{SU}(2)$. Such coadjoint orbits are naturally Kähler and therefore admit quantization in terms of holomorphic variables -- a fact already evident from the form of the action \eqref{ActionSphere}.

Let us discuss such quantization in detail. From the symplectic structure \eqref{SymplecticFormSphere} we can read off the quantization rules similar to \eqref{FlatQuantizationRules},
\begin{equation}\label{QuantRulesSphere}
z^\alpha\mapsto \frac{\partial}{\partial z^\alpha}\,,\quad \smallthickbar{z}_{\,\alpha}\mapsto z^\alpha\,,\qquad w^\alpha\mapsto\frac{\partial}{\partial w^\alpha}\,,\quad \smallthickbar{w}_\alpha\mapsto w^\alpha\,.
\end{equation}
The space of states consists of holomorphic functions $\Phi(z,w)$ in $z^\alpha$ and $w^\alpha$, constrained to satisfy  
\begin{equation}\label{SphereConstr}
z^\alpha\frac{\partial\Phi}{\partial z^\alpha} = p\,,\qquad w^\alpha\frac{\partial\Phi}{\partial w^\alpha} = p+\mathfrak{q}\,,
\end{equation}
where the sum over $\alpha=1,2$ is implied. In more geometric terms, $\Phi$ is a holomorphic section of $\mathscr{O}(p)\boxtimes\mathscr{O}(p+\mathfrak{q})$ bundle over the product of spheres. 

The quantum Hamiltonian is 
\begin{equation}\label{SphereHamilt}
\widehat{H}_{\mathrm{spin}} = \bigg(z^\alpha\frac{\partial}{\partial w^\alpha}\bigg)\bigg(w^\beta\frac{\partial}{\partial z^\beta}\bigg)\,.
\end{equation}
We can also equip the space of our states with a $\mathbf{SU}(2)$-invariant Hermitian scalar product, and it can be checked that this Hamiltonian is self-adjoint with respect to this product, see Appendix \ref{NormAppendix}.

The formal solutions of the eigenvalue problem $\widehat{H}_{\mathrm{spin}}\Phi(z,w) = E\Phi(z,w)$ are given by 
\begin{equation}\label{SphereFormalStates}
\Phi(z,w|\upgamma) = \big(\varepsilon_{\alpha\beta}z^\alpha w^\beta\big)^l\big(z\circ\upgamma\big)^n\big(w\circ\upgamma\big)^m\,,\qquad E=n(m+1)\,,
\end{equation}
where $\upgamma$ is a fixed reference vector in $\mathbb{C}^2$ and $z$ and $w$ are the homogeneous coordinates. Our goal here is to determine for which values of $l$, $n$, and $m$ such formal solutions represent well-defined elements of our Hilbert space.

At the very least, they must correspond to single-valued functions. This implies that if an entire function has a zero, then to avoid branch cuts we can only take integer powers of that function. For example, $z\circ\upgamma$ has a zero when $z^\alpha = \varepsilon^{\alpha\beta}\upgamma_\beta$, so $n$ must be an integer. For similar reasons, we fix $l,n,m\in\mathbb{Z}$. The constraints \eqref{SphereConstr} imply that
\begin{align}
l+n = p\,,\qquad l+m = p+\mathfrak{q}\,,\qquad\Longrightarrow\qquad m = n+\mathfrak{q}\,,
\end{align}
which also impose integrality conditions on $p$ and $\mathfrak{q}$. Furthermore, holomorphic functions must not contain any poles, so $l$, $m$, and $n$ must also be non-negative. Summarizing this analysis, we find that the functions
\begin{equation}\label{WellDefinedSectionsSphere}
\Phi(z,w|\upgamma) = \big(\varepsilon_{\alpha\beta}z^\alpha w^\beta\big)^{p-n}\big(z\circ\upgamma\big)^n\big(w\circ\upgamma\big)^{n+\mathfrak{q}}\,,
\end{equation}
where
\begin{equation}\label{spherespectrum}
E=n(n+\mathfrak{q}+1)\,,\quad n = 0,\ldots,p\,,
\end{equation}
are well-defined elements of the Hilbert space. Thus, we recover in the limit $p\rightarrow\infty$ the correct spectrum of a particle on the sphere in the presence of a magnetic monopole of charge~$\mathfrak{q}$. The ordinary spherical harmonics on the sphere also can be recovered from the holomorphic sections \eqref{WellDefinedSectionsSphere}, see the discussion below, Appendix \ref{AppendixFlatRelation} and \cite{Bykov:2023uwb,Krivorol2026}. The  sections \eqref{WellDefinedSectionsSphere} are also the irreducible $\mathbf{SU}(2)$ representations of spin $n+\frac{\mathfrak{q}}{2}$ under the diagonal group action on the $z$ and $w$ variables \cite{Bykov:2024tvb}. 

Note that the eigenfunctions \eqref{WellDefinedSectionsSphere} are also homogeneous of degree $2n+\mathfrak{q}$ with respect to the rescaling of $\upgamma$  and thus can be considered as holomorphic sections of $\mathscr{O}(2n+\mathfrak{q})$. The dichotomy between $z,w$ and $\upgamma$ variables manifests the fact that the eigenfunctions $\Phi(z,w|\upgamma)$ `intertwine' $\mathscr{O}$-bundles over $\mathbb{CP}^1\times\mathbb{CP}^1$ with bundles over a single copy of $\mathbb{CP}^1$. It also explains the fact that $\Phi(z,w|\upgamma)$ carry the spin $n+\frac{\mathfrak{q}}{2}$. They can also be seen as $\mathbf{SU}(2)$ coherent states \cite{PerelomovBook}, because they can be obtained from a reference state, say
\begin{equation}
\Phi_{\mathsf{ref}}(z,w) = \big(\varepsilon_{\alpha\beta}z^\alpha w^\beta\big)^{p-n}\big(z^1w^1\big)^n\big(w^1\big)^{\mathfrak{q}}\,,
\end{equation}
by the diagonal $\mathbf{SU}(2)$ action.

It is also possible to represent the biholomorphic states of spin \(n+\frac{\mathfrak{q}}{2}\) using the space of totally symmetric tensors of rank \(2n+\mathfrak{q}\), cf.~\cite{Bykov:2024tvb}. To achieve this, consider a superposition of the `coherent states' \eqref{WellDefinedSectionsSphere} of the form 
\begin{align}
\Phi_{\mathsf{T}}(z,w) =& \int\dif^4\upgamma \,\mathsf{T}_{\alpha_1\ldots \alpha_{2n+\mathfrak{q}}}\smallthickbar{\upgamma}^{\alpha_1}\ldots\smallthickbar{\upgamma}^{\alpha_{2n+\mathfrak{q}}}\,\Phi(z,w|\upgamma)\,e^{-|\upgamma_1|^2-|\upgamma_2|^2}\sim\nonumber\\
\sim&\big(\varepsilon_{\alpha\beta}z^\alpha w^\beta\big)^{p-n}\,\mathsf{T}_{\alpha_1\ldots \alpha_{2n+\mathfrak{q}}}z^{\alpha_1}\ldots z^{\alpha_n} w^{\alpha_{n+1}}\ldots w^{\alpha_{2n+\mathfrak{q}}}\,,\label{Ttransformation}
\end{align}
where we apply Wick's theorem for Gaussian integrals and omit a combinatorial coefficient in the second line. Here \(\mathsf{T}\) is a symmetric tensor that defines the spin \(n+\frac{\mathfrak{q}}{2}\) representation of \(\mathbf{SU}(2)\). Note that the transformation~\eqref{Ttransformation} is a close analogue of~\eqref{FlatMagneticTransformation}.

\subsubsection{Flat space limit.} The flat space action can be obtained by taking the limit of infinite radius of the sphere in the sigma model. This limit can be performed explicitly by switching to inhomogeneous coordinates on the product of the spheres. In what follows, we work with the inhomogeneous parametrization $w=[w^1:w^2]\mapsto [1:w]$, $z=[z^1:z^2]\mapsto[-z:1]$ and $\upgamma=[\upgamma^1:\upgamma^2]\mapsto[1:\gamma]$. In these coordinates, the action~\eqref{ActionSphere} for a sphere of radius $R$ takes the form
\begin{equation}
\mathcal{S} = \int\dif t\,\Bigg(\frac{ip\big(\smallthickbar{z}\dot{z} - z\dot{\smallthickbar{z}}\big)}{2\big(1+|z|^2\big)}
+\frac{i(p+\mathfrak{q})\big(\smallthickbar{w}\dot{w} - w\dot{\smallthickbar{w}}\big)}{2\big(1+|w|^2\big)} -  \frac{p(p+\mathfrak{q})}{R^2}\frac{|z-\smallthickbar{w}|^2}{(1+|z|^2)(1+|w|^2)}\Bigg)\,.
\end{equation}
The flat space action \eqref{FlatSpinChainAction} is recovered by rescaling the variables as $z\rightarrow R^{-1}z$ and $w\rightarrow R^{-1} w$ and taking the limit $R\rightarrow\infty$, while keeping $\lambda:=R^{-2}p$ and $B:=R^{-2}\mathfrak{q}$  fixed. This clarifies the origin of the auxiliary parameter $\lambda$ introduced in the earlier sections.

Let us now see how the eigenfunctions (\ref{WellDefinedSectionsSphere}) approach the ones of the Landau problem in flat space~(\ref{PhiNstate}) in the same limit. First we express these eigenfunctions using inhomogeneous coordinates: 
\begin{align}
    \Phi(z,w|\upgamma) \sim (1+zw)^{p-n} \left(1-{z\over \gamma}\right)^n (1+w\gamma)^{n+\mathfrak{q}}\,,
\end{align}
where $z$, $w$ and $\gamma$ are the inhomogeneous coordinates.
In order to take the limit, we set $p=\lambda R^2$,  $\mathfrak{q}=B R^2$ and rescale $(z, w, \gamma)\to {1\over R} (z, w, \gamma)$. Sending $R\to\infty$, one then obtains
\begin{align}
    \Phi(z,w|\upgamma) \sim e^{\lambda z w+B  \gamma w} \left(1-{z\over \gamma}\right)^n\,,\quad\quad n=0, 1,  \ldots, \infty\,,
\end{align}
reproducing~(\ref{PhiNstate}) up to the redefinition $\gamma={\beta\over B}$. In this limit the energy spectrum~(\ref{spherespectrum}) also has the correct asymptotic behavior
\begin{align}
    {E\over R^2}\mapsto n B\,,\quad\quad n=0, 1,  \ldots, \infty\,,
\end{align}
coinciding with~(\ref{Landauenergy}).

\section{The hyperbolic plane $\mathbb{H}$ and $\mathbf{SU}(1, 1)$ orbits}\label{H2sec}
In the previous sections, we applied the holomorphic quantization method to analyze particles on two-dimensional Riemannian manifolds with constant zero or positive curvature, that is, on a plane and on a sphere. In this section, we consider the final case of a particle on the hyperbolic plane. This case incorporates the features of the previous two cases, including the non-compactness of the configuration space and its symmetry group, and the presence of both discrete and continuous spectra in a transverse magnetic field, see
\cite{ComtetLandauHyperbolic,DUNNE1992233,Grosche:1987de,Grosche:1988um,Balazs:1986uj,Kuperin:1994eh,Kim:2001tw}. 

Our strategy follows the same principle: we establish a symplectomorphism between the particle’s phase space and the product of two copies of its configuration space, each of which is a coadjoint orbit of the symmetry group $\mathbf{SU}(1,1) \simeq \mathbf{SL}(2,\mathbb{R})$. These two coadjoint orbits, namely the hyperbolic planes, are Kähler manifolds and therefore admit a natural holomorphic quantization.

\subsection{Symplectic structure.} 
The Lobachevsky (or hyperbolic) plane $\mathbb{H}$ can be modeled on the unit disc $\mathbb{D}\subset\mathbb{C}$ endowed with the metric
\begin{equation}\label{Hyperbolic2Metric}
\dif s^2 = \frac{\dif\xi\dif\smallthickbar{\xi}}{\big(1-|\xi|^2\big)^2}\,,\qquad |\xi|<1\,.
\end{equation}
The hyperbolic plane also can be seen as the coset space
\begin{equation}\label{HasCoset}
\mathbb{H}\simeq \frac{\mathbf{SU}(1,1)}{\mathbf{U}(1)}\,,
\end{equation}
where $\mathbf{SU}(1,1)$ is the group of  matrices $\mathbf{g}$ of the form
\begin{equation}\label{SU11mat}
\mathbf{g}=
\begin{pmatrix}
\alpha & \beta \\
\smallthickbar{\beta} & \smallthickbar{\alpha}
\end{pmatrix}\,,\qquad |\alpha|^2 - |\beta|^2 = 1\,,\qquad\alpha,\beta\in\mathbb{C}\,.
\end{equation}
In fact, $\mathbb{H}$ is a (co)adjoint orbit of $\mathbf{SU}(1,1)$ and thus it is a symplectic manifold. 

For further use we introduce the indefinite norm in $\mathbb{C}^2$  as follows:
\begin{align}\label{indefinitenorm}
    \lVert v\rVert^2 =\smallthickbar{v}^{\,\alpha}\eta_{\alpha\beta} v^\beta= |v_1|^2-|v_2|^2\,,\qquad \eta = \begin{pmatrix}
1 & 0 \\
0 & -1
\end{pmatrix}\,.
\end{align}

We will be dealing with two slightly different parameterizations of the symplectic structure. To describe the first one, we define a complex vector $z\in\mathbb{C}^2$ such that ${\lVert z \rVert^2=1}$, and we identify $z\sim e^{i\varphi}z$ ($\varphi\in\mathbb{R}$). 
Such vectors clearly parametrize~\eqref{HasCoset} as the coset space. The $\mathbf{SU}(1,1)$ invariant symplectic form in this parameterization has the structure
\begin{equation}
\omega_p^+ = ip\,\dif\smallthickbar{z}\wedge\eta\dif z := ip\big(\dif\smallthickbar{z}_{\,1}\wedge \dif z^1 - \dif\smallthickbar{z}_{\,2}\wedge \dif z^2\big)\,,\qquad p\in\mathbb{R}\,.
\end{equation}
We denote the parameterization $\big(\mathbb{H},\omega_p^+\big)$ as $\mathbb{H}^+_p$.
In the second parameterization we use the complex vector $w\in\mathbb{C}^2$ satisfying  ${\lVert w \rVert^2=-1}$ and subject to the identification $w\sim e^{i\varphi}w$. 
The invariant symplectic form is
\begin{equation}
\omega_p^- = -ip\,\dif\smallthickbar{w}\wedge\eta\dif w := -ip\big(\dif\smallthickbar{w}_{\,1}\wedge \dif w^1 - \dif\smallthickbar{w}_{\,2}\wedge \dif w^2\big)\,.
\end{equation}
We denote the parameterization $\big(\mathbb{H},\omega_p^-\big)$ as $\mathbb{H}^-_p$.

The manifolds $\mathbb{H}^+_p$ and $\mathbb{H}^-_p$ are clearly symplectomorphic (mapped into each other by $z^1\mapsto w^2, z^2\mapsto w^1$) but we equip them with inequivalent  $\mathbf{SU}(1,1)$-actions 
\begin{equation}
\begin{pmatrix}
z^1 \\ z^2     
\end{pmatrix}\rightarrow \begin{pmatrix}
\alpha & \beta \\
\smallthickbar{\beta} & \smallthickbar{\alpha}
\end{pmatrix}\begin{pmatrix}
z^1 \\ z^2     
\end{pmatrix}\,,\qquad
\begin{pmatrix}
w^1 \\ w^2     
\end{pmatrix}\rightarrow \begin{pmatrix}
\alpha & \beta \\
\smallthickbar{\beta} & \smallthickbar{\alpha}
\end{pmatrix}\begin{pmatrix}
w^1 \\ w^2     
\end{pmatrix}\,.
\end{equation}
In other words, this symplectomorphism is not equivariant w.r.t.~the group action. This explains why the geometric quantization of $\mathbb{H}^\pm_p$ is given by the inequivalent discrete series representations $\mathcal{D}^\pm_p$ of $\mathbf{SU}(1,1)$, see the discussion below.

\subsection{Classical aspects.} 

We are interested in a `classical spin chain' with the phase space $\mathbb{H}^+_p\times\mathbb{H}^-_p$ and a  Hamiltonian of the form 
\begin{equation}
H_{\text{spin}} = p^2|\smallthickbar{z}\eta w|^2 = p^2\big(\smallthickbar{z}_1w^1-\smallthickbar{z}_2w^2\big)\big(\smallthickbar{w}_1z^1-\smallthickbar{w}_2z^2\big)\,.
\end{equation}
The crucial property of this Hamiltonian is that it vanishes on the variety 
\begin{equation}
\Bigl\{\;\smallthickbar{z}\eta w = 0\;\Bigr\}\subset \mathbb{H}^+_p\times\mathbb{H}^-_p\,.
\end{equation}
This Hamiltonian can also be written in the ``inhomogeneous coordinates''
\begin{equation}
z:=\frac{z^2}{z^1}\,,\quad w := \frac{w^1}{w^2}\,,\qquad\Longrightarrow\qquad |z|^2\leq 1\,,\quad |w|^2\leq 1
\end{equation}
as
\begin{equation}
H_{\mathrm{spin}} = p^2\frac{|z - \smallthickbar{w}|^2}{\big(1-|z|^2\big)\big(1-|w|^2\big)}\,,
\end{equation}
where we used the normalizations of $z^\alpha$ and $w^\alpha$ vectors. It follows from the inequalities $|z|^2\leq 1$ and $|w|^2\leq 1$ that in the product of Lobachevsky planes such a Hamiltonian has a critical submanifold only on the `diagonal' $z = \smallthickbar{w}$ (which is of course a global minimum). Again, the above formula for the Hamiltonian should be compared with~(\ref{flatspaceHam}), (\ref{S2Ham}) for flat space and the $2$-sphere respectively. The minimal submanifold is a copy of $\mathbb{H}$, so that one has the embedding
\begin{align}\label{HyperLagrEmb}
    (\mathbb{H})_{\mathrm{min}}\hookrightarrow \mathbb{H}^+_p\times\mathbb{H}^-_p\,.
\end{align}
As in all the previous cases, this submanifold is Lagrangian in the absence of magnetic field and symplectic for $B\neq 0$. 

\subsection{The (indefinite signature) polar decomposition.} 

We claim that the `spin chain'  $\big(\mathbb{H}^+_p\times\mathbb{H}^-_p,H_{\text{spin}}\big)$ in fact describes a free particle on the hyperbolic plane. To prove this, we first write out the action for this classical spin chain in homogeneous coordinates:
\begin{equation}\label{FirstHyperbAction}
\mathcal{S}[z,w] = \int\dif t\,\Big[ip\big(\smallthickbar{z}\eta\dot{z} - \smallthickbar{w}\eta\dot{w}\big) - p^2|\smallthickbar{z}\eta w|^2\Big]\,,\qquad 
\lVert z\rVert^2 = 1\,,\quad \lVert w\rVert^2 = -1\,,
\end{equation}
where $\lVert\bullet\rVert$ is the indefinite norm in $\mathbb{C}^2$  introduced in~(\ref{indefinitenorm}).  
Denote by $\mathcal{Z} = \big(z~~w\big)$ the $2\times 2$ matrix with columns $z$ and $w$. 
Next, we perform the indefinite analog of the polar decomposition of the form~\cite{Bolshakov} 
\begin{equation}\label{IndPolarDecomposition}
\mathcal{Z} = U\sqrt{G}\,,\qquad U\in\mathbf{U}(1,1)\,.
\end{equation}
Since $U\in\mathbf{U}(1,1)$, it satisfies $U^\dagger \eta U=\eta$. Analogously, $G$ is an analogue of Hermitian matrix, satisfying $G^\dagger=\eta G \eta$. It then easily follows that 
\begin{equation}
G := \eta\mathcal{Z}^\dagger\eta \mathcal{Z} = \begin{pmatrix}
1 & \smallthickbar{z}\eta w\\
-\smallthickbar{w}\eta z & 1
\end{pmatrix}:= \begin{pmatrix}
1 & g\\
-\smallthickbar{g} & 1
\end{pmatrix}\,,
\end{equation}
where we used the fact that the $z$ and $w$ vectors are normalized. Note that $G$ is  positive-definite, $
\det(G) = 1+|g|^2>0
$, 
and thus we can extract the (positive-definite) square root $\sqrt{G}$. Moreover, the variable $g\in \mathbb{C}$ is unconstrained\footnote{This is what makes this case different from that of the sphere $\mathbb{S}^2$, where an analogous variable is only unconstrained in the limit $p\to \infty$; see the discussion in~\cite{Bykov:2024tvb,Krivorol2026}.}.  
The action \eqref{FirstHyperbAction} can be rewritten as
\begin{equation}\label{SpinChainMatrixAction}
\mathcal{S}[\mathcal{Z}] = \int\dif t\,\Big[ip\mathsf{Tr}\big( \eta\mathcal{Z}^\dagger\eta \dot{\mathcal{Z}}\big)-p^2 |g|^2\Big]\,.
\end{equation}
We also denote the column vectors of the matrix $U$ as $u$ and $v$. Since $\eta U^\dagger\eta U = \mathds{1}_2$, they are normalized as
\begin{equation}\label{HyperUVconstraints}
\smallthickbar{u}\eta u = 1\,,\qquad \smallthickbar{v}\eta v = -1\,,\qquad \smallthickbar{u}\eta v = \smallthickbar{v}\eta u =0\,.
\end{equation}

\subsection{Mapping to a sigma model.}

In order to demonstrate that the spin chain above reproduces particle dynamics on the hyperbolic plane, we will first rewrite the kinetic term 
\begin{equation}
\mathcal{S}_{\mathrm{kin}}[\mathcal{Z}] = \int\dif t\,\Big[ip\mathsf{Tr}\big( \eta\mathcal{Z}^\dagger\eta \dot{\mathcal{Z}}\big)\Big]
\end{equation}
in terms of the $U$ and $G$ matrices. By introducing the matrix $H:=\sqrt{G}$ and substituting the indefinite polar decomposition \eqref{IndPolarDecomposition}, we get
\begin{align}
\mathcal{S}_{\mathrm{kin}} =& ip\int\dif t\,\mathsf{Tr}\bigg(\eta H^\dagger U^\dagger\eta\,\frac{\dif}{\dif t}\big(UH\big)\bigg) =\nonumber\\
=&ip\int\dif t\,\mathsf{Tr}\Big(\underbrace{\big(\eta H^\dagger\eta\big)}_{=H}\underbrace{\big(\eta U^\dagger\eta\big)U}_{=\mathds{1}}\dot{H}+\underbrace{\big(\eta H^\dagger\eta\big)}_{=H}\big(\eta U^\dagger\eta\big)\dot{U}H\Big)\,,
\end{align}
where we inserted the unit matrix $\mathds{1}=\eta^2$ in the second line. The first term in the second line vanishes identically, since it is a derivative of a constant, i.e.
\begin{equation}
\frac{\dif}{\dif t}\mathsf{Tr}\big(H\dot{H}\big) = \frac{1}{2}\frac{\dif}{\dif t}\underbrace{\mathsf{Tr}(G)}_{=2} = 0\,.
\end{equation}
Thus, the kinetic term can be rewritten as
\begin{equation}
\mathcal{S}_{\mathrm{kin}} = ip\int\dif t\,\mathsf{Tr}\big(G\,\eta U^\dagger\eta\dot{U}\big)\,.
\end{equation}
By substituting the parameterization of $U$ in terms of $u$ and $v$ and taking the trace, we get
\begin{equation}
\mathcal{S}_{\mathrm{kin}} = ip\int\dif t\big(\smallthickbar{u}\eta\dot{u}-\smallthickbar{v}\eta\dot{v}+\smallthickbar{g}\,\smallthickbar{u}\eta\dot{v}+g\,\smallthickbar{v}\eta\dot{u}\big)\,.
\end{equation}
The term $\smallthickbar{u}\eta\dot{u}-\smallthickbar{v}\eta\dot{v}$ is a total derivative. Geometrically, this follows because it arises as the pullback to the worldline of the Liouville $1$-form whose exterior derivative yields a vanishing symplectic form:
\begin{equation}
\omega = i\dif\smallthickbar{u}\wedge\eta\dif u-i\dif\smallthickbar{v}\wedge\eta\dif v = 0\,,
\end{equation}
see the parameterization \eqref{HyperUVconstraints} and recall the Lagrangian embedding \eqref{HyperLagrEmb}.
Thus, after the rescaling $g\rightarrow p^{-1}g$, the spin chain action (\ref{SpinChainMatrixAction}) can be rewritten in the form
\begin{equation}\label{guv action}
\mathcal{S}[u,v,g] = \int\dif t\,\big(i\smallthickbar{g}\,\smallthickbar{u}\eta\dot{v}+ig\,\smallthickbar{v}\eta\dot{u}-|g|^2\big)\,.
\end{equation}
Eliminating $g$ and $\smallthickbar{g}$ by their equations of motion, we get an equivalent action
\begin{equation}
\mathcal{S}[u,v] = \int\dif t\,|\smallthickbar{u}\eta\dot{v}|^2\,.
\end{equation}
We can finally check that this is the sigma model on the unit disc $\mathbb{D}$ with the metric~(\ref{Hyperbolic2Metric}) by taking an explicit parameterization of (\ref{HyperUVconstraints}),
\begin{equation}
u = \frac{1}{\sqrt{1-|\xi|^2}}\begin{pmatrix}
1 \\ \smallthickbar{\xi}
\end{pmatrix}\,,\qquad v = \frac{1}{\sqrt{1-|\xi|^2}}\begin{pmatrix}
\xi \\ 1
\end{pmatrix}\,,\qquad\xi\in\mathbb{D}\,.
\end{equation}

The generalized polar decomposition~(\ref{IndPolarDecomposition}) is a map from $\mathbb{H}^+_p\times\mathbb{H}^-_p$ parametrized by $(
    z \;, w
)\sim (z\, e^{i\phi}, w \, e^{i\psi})$ to the cotangent bundle $\mathrm{T}^\ast(\mathbb{H})_{\mathrm{min}}$ parametrized by the triple $(u, v, g)\sim (u\, e^{i\phi}, v \, e^{i\psi}, g\,e^{i(\psi-\phi)})$, where $u$ and $v$ are subject to~(\ref{HyperUVconstraints}). Since $g$ is unconstrained, this is in fact a symplectomorphism\footnote{This phase space -- the product of two coadjoint orbits of $\mathbf{SL}(2,\mathbb{R})$ --  also arises in the problem of quantization of a  relativistic particle on the $\mathbf{SL}(2,\mathbb{R})$ group manifold \cite{Dzhordzhadze:1994np}.} 
\begin{align}
\mathrm{T}^\ast(\mathbb{H})_{\mathrm{min}}\simeq \mathbb{H}^+_p\times\mathbb{H}^-_p\,.
\end{align}
Indeed, the kinetic term in~(\ref{guv action}) is the pull-back of the Poincar\'e-Liouville one-form of the cotangent bundle, written using invariant variables.

\subsection{Orbits of $\mathbf{SU}(1, 1)$.} In the case of $\mathfrak{su}(2)$ every element of the Lie algebra is conjugate to an element of the Cartan subalgebra, i.e.~to a traceless diagonal matrix. This is simply because every Hermitian matrix can be diagonalized. As  a result, the only (co)adjoint orbits (apart from the trivial one) in that case are spheres ${\mathbf{SU}(2)\over \mathbf{U}(1)}=\mathbb{S}^2$.

However, this is no longer the case for $\mathfrak{su}(1,1)$. This algebra can be abstractly realized as \cite{PerelomovBook}
\begin{equation}
[K_1,K_2] = -K_0\,,\qquad [K_0,K_1] = K_2\,,\qquad [K_2,K_0] = K_1\,.
\end{equation}
The coadjoint orbits, as in Section \ref{SectionISOcoadjointOrbits}, can be computed as symplectic leaves in the dual space $\mathfrak{su}^\ast(1,1)$ equipped with the  Kirillov-Kostant-Souriau Poisson bracket. It can be identified with the Minkowski space $\mathbb{R}^{2,1}$ with the coordinates $x_0, x_1, x_2$ and the Poisson structure
\begin{equation}
\{x_1,x_2\} = -x_0\,,\qquad \{x_0,x_1\} = x_2\,,\qquad \{x_2,x_0\} = x_1\,.
\end{equation}
There is the single Casimir function of the form $\mathcal{C} = x_0^2-x_1^2-x_2^2$ which Poisson commutes with everything. Thus, possible $\mathfrak{su}(1,1)$ coadjoint orbits are classified by the level surfaces of $\mathcal{C}$.
There are three cases (cf.~\cite{WittenVirasoro, BasileOrbits}): 
\begin{align}
    &x_0^2-x_1^2-x_2^2=-R^2< 0\quad\Rightarrow\quad \textrm{One-sheeted hyperboloid}\,,\\
    &x_0^2-x_1^2-x_2^2=R^2> 0\quad\Rightarrow\quad \textrm{Two-sheeted hyperboloid}\,,\\
    &x_0^2-x_1^2-x_2^2=0\quad\Rightarrow\quad \textrm{Cone}\,.
\end{align}
Leaving aside the special case of the cone, the other two cases correspond to two types of unitary representations of $\mathfrak{su}(1,1)$. The one-sheeted hyperboloid $\mathbf{SO}(1,2)/\mathbf{SO}(1,1)$ may be viewed as $\mathrm{T}^\ast \mathbb{S}^1$, and its quantization gives the continuous series representations $\mathcal{C}_q$ of $\mathfrak{su}(1,1)\simeq\mathfrak{sl}_2(\mathbb{R})$, see \cite{KitaevSL2R,Plyushchay1993,Neri:2025fsh}. In the case of the two-sheeted hyperboloid each sheet is a copy of ${\mathbf{SO}(1,2)\over \mathbf{SO}(2)}\simeq \mathbb{H}_2$, the Lobachevsky plane. Its quantization brings about representations of the discrete series, denoted $\mathcal{D}_p^{\pm}$ ($\pm$ depending on the sheet). The latter are representations with a lowest/highest weight, whereas continuous series representations have no lowest or highest weight.

\subsection{Holomorphic quantization.}
Here we want to quantize the particle on the hyperbolic plane in a manner similar to the  case of $\mathbb{S}^2$. Let us also introduce the magnetic field in the model similarly to \eqref{ActionSphere}, i.e.~our action has the form
\begin{equation}
\mathcal{S}[z,w] = \int\dif t\,\Big[i\big(\smallthickbar{z}\eta\dot{z} - \smallthickbar{w}\eta\dot{w}\big) - |\smallthickbar{z}\eta w|^2\Big]\,,\qquad 
\lVert z\rVert^2 = p\,,\quad \lVert w\rVert^2 = -(p+\mathfrak{q})\,,
\end{equation}
where $\mathfrak{q}$ is the monopole charge. Note that the structure of 
this action is the same as of~\eqref{ActionSphere} except the presence of the indefinite metric $\eta$, which leads to the  modified quantization rules \eqref{QuantRulesSphere} for homogeneous coordinates:
\begin{equation}
z\mapsto \frac{\partial}{\partial z}\,,\quad \smallthickbar{z}\eta\mapsto z\,,\qquad w\mapsto\frac{\partial}{\partial w}\,,\quad -\smallthickbar{w}\eta\mapsto w\,.
\end{equation}
In this way we absorb the metric $\eta$ into the definition of the quantization map and thus the holomorphic quantization is almost identical to the case of Section~\ref{SphereHolomorphicQuant}.
Namely, the space of states consists of holomorphic functions $\Phi(z,w)$ in the vectors $z,w$ satisfying the constraints \eqref{SphereConstr},
\begin{equation}\label{HyperbConstraints}
z^\alpha\frac{\partial\Phi}{\partial z^\alpha} = -p:=-p_1\,,\qquad w^\alpha\frac{\partial\Phi}{\partial w^\alpha} = -(p+\mathfrak{q}):=-p_2\,,
\end{equation}
where we have additionally reversed the signs in the r.h.s.\footnote{This sign in our definition of the states is a matter of convention. In practice, the allowed values of $p_{1,2}$ are defined by the normalizability of states, see the discussion below. As we will see, this extra sign is natural in the hyperbolic case.}. Holomorphicity implies that $p_{1,2}\in\mathbb{Z}$. 
The Hamiltonian takes the form \eqref{SphereHamilt} with the extra minus sign\footnote{As will be clear in what follows, and as we explain in Appendix~\ref{NormAppendix}, due to this extra minus sign the Hamiltonian is positive-definite.},
\begin{equation}\label{SpinHyperbolicHamiltonian}
\widehat{H}_{\mathrm{spin}} = -\bigg(z^\alpha\frac{\partial}{\partial w^\alpha}\bigg)\bigg(w^\beta\frac{\partial}{\partial z^\beta}\bigg)\,.
\end{equation}
Thus, the formal wavefunctions still have the form \eqref{SphereFormalStates}, i.e.
\begin{equation}\label{FormalWavesHyperbolic}
\Phi(z,w|\upgamma) = \big(\varepsilon_{\alpha\beta}z^\alpha w^\beta\big)^l\big(z\circ\upgamma\big)^n\big(w\circ\upgamma\big)^m\,,\qquad E=-n(m+1)\,,
\end{equation}
where now $\circ$ means the product in $\mathbb{C}^2$ in the indefinite metric $\eta$.
The constraints \eqref{HyperbConstraints} imply
\begin{align}\label{lnm constraints}
    l+n=-p_1\,,\quad\quad l+m=-p_2\,.
\end{align}

\begin{figure}
    \centering
    \begin{overpic}[width=0.7\linewidth]{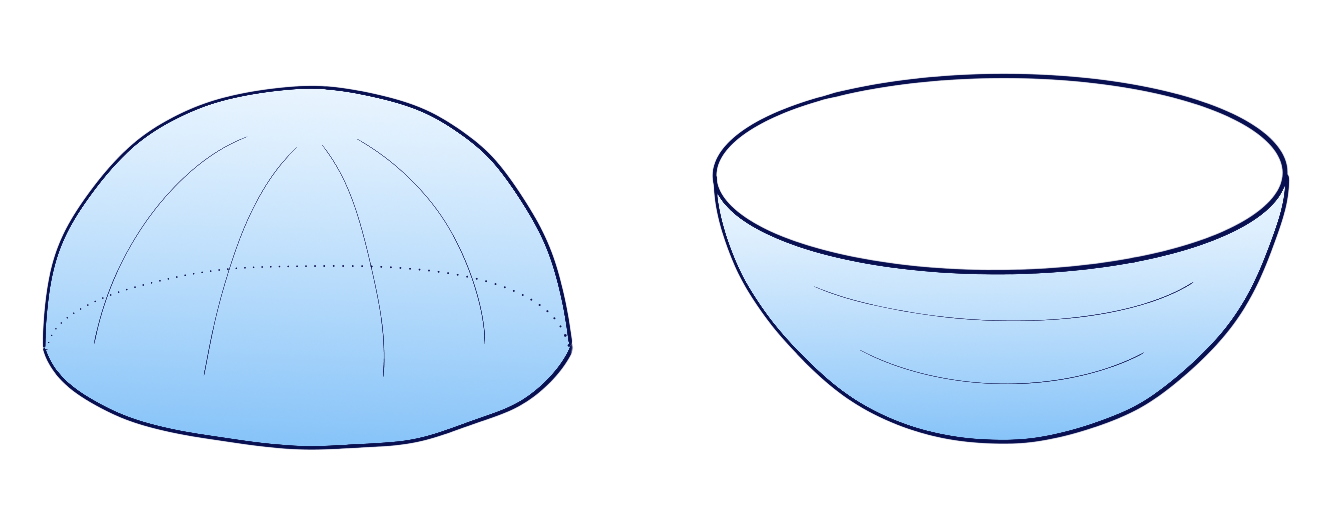}
    \put(143,50){$\times$}
    \put(-5,90){$\mathbb{H}_{p}^+$}
    \put(300,90){$\mathbb{H}_{p+\mathfrak{q}}^-$}
\end{overpic}
    \caption{The product of two copies of the Lobachevsky plane $\mathbb{H}_{p}^+ \times \mathbb{H}_{p+\mathfrak{q}}^-$, viewed as hemispheres of $\mathbb{CP}^1$ determined by $\lVert z\rVert^2>0$, $\lVert w\rVert^2<0$.}
    \label{UpDownSpheres}
\end{figure}

However, there is one significant difference that affects the transition from formal to physical wavefunctions. This difference lies in the domain of the holomorphic variables. 
Recall that
$\mathbb{H}_{p}^+$ and $\mathbb{H}_{p+\mathfrak{q}}^-$ can be viewed as submanifolds of $\mathbb{CP}^1$, defined by projective vectors $v\in\mathbb{P}\big(\mathbb{C}^2\big)$ with  $\lVert v\rVert^2>0$ and $\lVert v\rVert^2<0$, respectively, where $\lVert \bullet\rVert$ is the indefinite norm~(\ref{indefinitenorm}) (see Fig.~\ref{UpDownSpheres}). In other words, we use the `hemisphere models' for $\mathbb{H}^\pm$ and thus additionally impose the conditions 
\begin{align}
    \lVert z\rVert^2>0\quad\quad \textrm{and}\quad\quad \lVert w\rVert^2<0\,.
\end{align}
In this framework, quantum states correspond to sections of the Hermitian bundle $\mathscr{O}(-p) \boxtimes \mathscr{O}(-p-\mathfrak{q})$ pulled back to $\mathbb{H}_{p}^+ \times \mathbb{H}_{p+\mathfrak{q}}^-$.

To get a Hilbert space, we have to additionally define a scalar product on our space of functions. This is done in Appendix~\ref{NormAppendix} in terms of functions of homogeneous variables. However, for practical purposes we prefer to state the result here in terms of inhomogeneous variables taking values inside the unit disc:
\begin{align}\label{scalp1p2}
    (\Phi_2, \Phi_1):=\int\limits_{|z|<1,\, |w|<1}\,\mathrm{vol}_{p_1, p_2}\,\smallthickbar{\Phi}_2\,\Phi_1\,,
\end{align}
where here and below we will use the following shorthand notation for the measure:
\begin{equation}\label{volp1p2}
\mathrm{vol}_{p_1, p_2}:=\frac{\dif^2z\,\dif^2w}{(1-|z|^2)^{2-p_1}(1-|w|^2)^{2-p_2}}\,.
\end{equation}

\subsection{The spectrum on $\mathbb{H}$.}

The energy spectrum of a particle on $\mathbb{H}$ in a homogeneous magnetic field was first found in~\cite{ComtetLandauHyperbolic} (see also the review~\cite{Balazs:1986uj} for a detailed discussion of the zero magnetic field case). It was found that for non-zero magnetic field $\mathfrak{q}\neq 0$ there is a finite number of discrete energy levels (whose number is related to $\mathfrak{q}$) apart from the continuous spectrum.  The fact that the spectrum has both discrete and continuous parts, has a nice classical counterpart: some of the magnetic  geodesics on~$\mathbb{H}$ are compact (these are analogs of the circular orbits in flat space) whereas others are infinite, stretching from the boundary back to the boundary. As we shall see below, this is correlated with the fact that the tensor product  $\mathcal{D}^+_{p_1}\otimes\mathcal{D}^-_{p_2}$ with $p_2>p_1$ contains $\bigoplus\mathcal{D}^-_{n}$, where the sum is over a finite range of $n$ related to~$\mathfrak{q}$. There is also some resemblance to the behavior observed for $\mathscr{B}^{\lambda}(\mathbb{C})\otimes\mathscr{B}^{\lambda+B}(\mathbb{C})$ in section \ref{SubsectionFlatHolomorphicQuant}, where the spectrum is discrete when $B \neq 0$ and continuous when $B = 0$.

To recover these results within our framework, we need to analyze when the formal wavefunctions \eqref{FormalWavesHyperbolic} correspond to  well-defined states in our Hilbert space.  
Thus, let us discuss the analytic structure of the eigenfunctions \eqref{FormalWavesHyperbolic}.

\subsubsection{Simplified case: $\mathcal{D}^+_p\otimes\mathcal{D}^+_{p+\mathfrak{q}}$.} 
\begin{figure}
    \centering
    
    \begin{overpic}[width=0.7\linewidth]{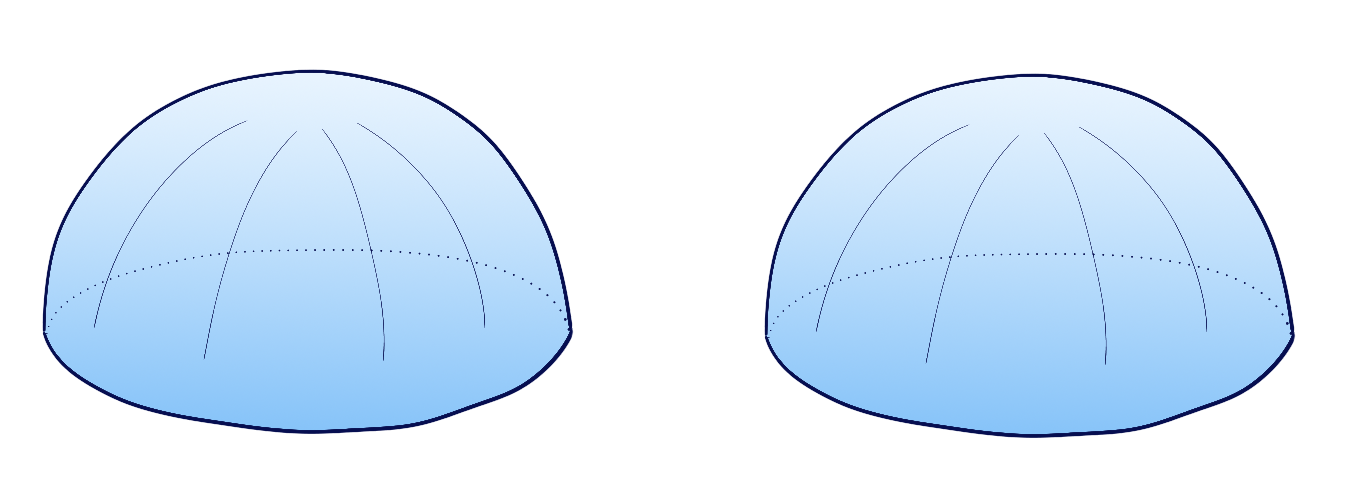}
    \put(147,50){$\times$}
    \put(-5,90){$\mathbb{H}_{p}^+$}
    \put(295,90){$\mathbb{H}_{p+\mathfrak{q}}^+$}
    
    \end{overpic}
    \caption{The product of two copies of the Lobachevsky plane $\mathbb{H}_{p}^+\times\mathbb{H}_{p+\mathfrak{q}}^+$, viewed as hemispheres of $\mathbb{CP}^1$ determined by $\lVert z\rVert^2>0$, $\lVert w\rVert^2>0$.}
    \label{UpUpSpheres}
\end{figure}

Let us first make a slight digression and analyze the simplified case when we quantize the space $\mathbb{H}_{p}^+\times\mathbb{H}_{p+\mathfrak{q}}^+$, obtaining the Hilbert space $\mathcal{D}^+_p\otimes\mathcal{D}^+_{p+\mathfrak{q}}$. In this case we should deal with the functions \eqref{FormalWavesHyperbolic} restricted to $\lVert z \rVert^2>0$ and $\lVert w \rVert^2>0$ (see Fig.~\ref{UpUpSpheres}). 

Here and hereafter it is more convenient to work in inhomogeneous coordinates. Let us introduce them as $w=[w^1:w^2]\mapsto [1:w]$ and $z=[z^1:z^2]\mapsto[1:z]$, where for now $|z|<1$ and $|w|<1$. For the vector $\upgamma$ we also introduce the inhomogeneous coordinate as $\upgamma = [\gamma,1]$, where $|\gamma|>1$ if $\lVert \upgamma \rVert^2>0$ (``spacelike''), $|\gamma|<1$ if $\lVert \upgamma \rVert^2<0$ (``timelike'') and $|\gamma|=1$ if $\lVert \upgamma \rVert^2=0$ (``lightlike''). In fact, here we will show that wavefunctions \eqref{FormalWavesHyperbolic} are normalizable only when $\upgamma$ is ``spacelike''. 

We start from the most interesting case $\lVert \upgamma \rVert^2>0$.
Here the wave function in inhomogeneous coordinates has the form
\begin{align}\label{D+D+waves}
    \Phi(z,w|\gamma)=(z-w)^l (z-\gamma)^n (w-\gamma)^m\,,\quad\quad |\gamma|>1\,.
\end{align}
Since $z-w$ has a zero, by the above arguments we assume $l\in \mathbb{Z}$, and therefore $n, m\in \mathbb{Z}$ as well due to~(\ref{lnm constraints}). We wish to find out for which values of the parameters the norm $\lVert \Phi \rVert^2=(\Phi, \Phi)$, where the scalar product is given by~(\ref{scalp1p2}), is finite:
\begin{align}
    \lVert \Phi \rVert^2=\int\,\mathrm{vol}_{p_1, p_2}
    \,|z-w|^{2l}\,|z-\gamma|^{2n}\,|w-\gamma|^{2m}\,.\label{ConvergentNormIntegralDplusDplus}
\end{align}
Since $|\gamma|>1$, the only potential singularity of the integrand is at $z=w$. Convergence of the integral then requires $l\geq 0$ and $p_{1,2}\geq 2$. 
Due to the constraints $l+n=-p_1$ and $l+m=-p_2$, one calculates the energy
\begin{align}
    E=-n(m+1)=(p_1+l)(p_2+l-1)\,,\quad\quad l=0, 1, 2, \ldots
\end{align}

Note that, if we return to homogeneous coordinates, the function \eqref{FormalWavesHyperbolic}, considered as a function of $\upgamma$, is homogeneous of degree $n+m=-p_1-p_2-2l$ and thus corresponds to the representation $\mathcal{D}^+_{p_1+p_2+2l}$. This reproduces the well-known tensor product decomposition 
\begin{equation}\label{TensorProductDplusDplus}
\mathcal{D}^+_{p_1}\otimes\mathcal{D}^+_{p_1}\simeq\bigoplus_{l=0}^\infty \mathcal{D}^+_{p_1+p_2+2l}\,,
\end{equation}
see \cite[section 7]{Repka}.

It is now clear why the cases $\lVert \upgamma \rVert^2<0$ and $\lVert \upgamma \rVert^2=0$ are inconsistent: the wavefunctions \eqref{FormalWavesHyperbolic} are never normalizable. 
In these cases, additional potential singularities arise when $z = \gamma$ and $w = \gamma$ for a ``timelike'' $\upgamma$, and as $z\rightarrow\gamma$ and $w\rightarrow\gamma$ for a ``lightlike'' $\upgamma$. 
One can check that, due to the constraints $n=-p_1-l<0$ and $m=-p_2-l<0$ derived above, these singularities are not integrable. This explains why the tensor product decomposition \eqref{TensorProductDplusDplus} does not contain  representations of the negative discrete series as well as of the continuous series.

\subsubsection{Discrete spectrum within $\mathcal{D}^+_{p}\otimes\mathcal{D}^-_{p+\mathfrak{q}}$.} Now we are in a position to analyze the case related to the hyperbolic plane sigma model, corresponding to the domain $\lVert z\rVert^2>0$ and $\lVert w\rVert^2<0$ in homogeneous coordinates.

When $p_1-p_2\neq 0$, the spectrum involves some discrete energy levels corresponding to $L^2$-normalizable wave functions, and in this Section we reproduce this part of the spectrum.
As we shall see in the next section, the case $\lVert \upgamma \rVert^2=0$ corresponds to the continuous spectrum. Assuming $\lVert \upgamma \rVert^2\neq 0$ for the moment,  the general wave functions~\eqref{FormalWavesHyperbolic} specialize to
\begin{align}
\Phi(z,w|\gamma)=(1-zw)^l (z-\gamma)^n \left(w-{1\over \gamma}\right)^m\,,\label{DiscreteHyperbolicWaves}
\end{align}
where we introduce the inhomogeneous coordinates as $w=[w^1:w^2]\mapsto [w:1]$, $z=[z^1:z^2]\mapsto[1:z]$, and $\upgamma = [\gamma,1]$. By the reasoning similar to that described above, if $p_2-p_1:=\mathfrak{q}>0$, in order to have normalizable eigenstates the vector $\upgamma$ should be ``timelike'', i.e.~$|\gamma|<1$. 

Since there is a zero at $z=\gamma$, we deduce that $l, n, m \in \mathbb{Z}$. Here there are two potential singularities: at $z=\gamma$ as well as at $zw=1$. Convergence of the integral for $\lVert \Phi \rVert^2$ at $z=\gamma$ requires that $n\geq 0$. Since $l+n=-p_1$, one has
\begin{align}\label{Abound1}
l<-p_1+1 \,.   
\end{align}
The behavior at ${zw=1}$ is the same as that of the test wave function ${\Phi_0=(1-zw)^l}$, so it suffices to compute its norm (here $(\bullet)_k$ is the Pochhammer symbol):
\begin{align}
    \lVert \Phi_0 \rVert^2=&\int_{\mathbb{H}\times \mathbb{H}}\,\mathrm{vol}_{p_1, p_2}\,|1-zw|^{2l}=\sum\limits_{k=0}^\infty\,\frac{(-l)_k^2}{(k!)^2}\int_{\mathbb{H}\times \mathbb{H}}\,\mathrm{vol}_{p_1, p_2}\,|z|^{2k}\,|w|^{2k}\sim\\&\sim \frac{1}{\big[(-l-1)!\big]^2}\sum\limits_{k=0}^\infty\,\frac{\big[(k-l-1)!\big]^2}{(k+p_1-1)!(k+p_2-1)!}\,.
\end{align}
At large $k$ the series behaves as $\sum\limits_{k=0}^\infty\,k^{-2l-p_1-p_2}$, so convergence requires that
\begin{align}\label{Abound2}
l>{1\over 2}(1-p_1-p_2)\,.
\end{align}
The two bounds~(\ref{Abound1}) and~(\ref{Abound2}) imply that 
\begin{align}\label{RangeOfM}
    l=-p_1-n\,,\quad\quad 0\leq n <{1\over 2}(p_2-p_1-1) = {1\over 2}(\mathfrak{q}-1) \,.
\end{align}
The corresponding eigenvalue is
\begin{align}
    E=-n(m+1)=n(p_2-p_1-n-1) = n(\mathfrak{q}-1-n)\,.
\end{align}
From the point of view of representations, this means that 
\begin{align}
    \mathcal{D}^+_{p}\otimes\mathcal{D}^-_{p+\mathfrak{q}} = \bigoplus_{0\leq n< {1\over 2}(\mathfrak{q}-1)}\,\mathcal{D}^-_{\mathfrak{q}-2n}+\ldots ,
\end{align}
where the rest of the decomposition corresponds to continuous spectrum. This reproduces the result \cite[Theorem 7.3]{Repka}.

\subsection{Continuous spectrum within $\mathcal{D}^+_p\otimes\mathcal{D}^-_{p}$.}
We now turn to the continuum part of the hyperbolic plane spectrum. Unlike the discrete part, it is present for any values of $p_1$ and $p_2$. For simplicity, we will elaborate the case of vanishing magnetic field, i.e. we set $p_1 = p_2 := p$; the general case $p_1 \neq p_2$ can be treated similarly. 

Following the reasoning similar to that in the previous sections, one finds that for $\lVert \upgamma\rVert^2\neq 0$ the wave functions~(\ref{FormalWavesHyperbolic}) are not normalizable and not even $\delta$-normalizable. Thus we try the case of ``lightlike'' $\upgamma$, i.e.~$|\gamma| = 1$. Parametrizing $\gamma=e^{-i\theta}$, we get\footnote{For completeness we note that, for general $p_1, p_2$, the corresponding wavefunction is \begin{align}
    \Phi_\chi(z,w)=(1-zw)^{\chi-{p_1+p_2\over 2}} (1-e^{i\theta}z)^{-{p_1-p_2\over 2}-\chi} \left(1-e^{-i\theta}w\right)^{-{p_2-p_1\over 2}-\chi}\,.
\end{align}}
\begin{align}\label{Phiwavefunc}
    \Phi_\chi(z,w|\theta)=(1-zw)^{\chi-p} \big(1-e^{i\theta}z\big)^{-\chi} \left(1-e^{-i\theta}w\right)^{-\chi}\,,\quad\quad \theta\in \mathbb{S}^1\,,
\end{align}
where $\chi \in \mathbb{C}$ is a so far undetermined complex number. Note that these states furnish a single $\mathbf{SU}(1, 1)$ orbit, which  is a circle\footnote{This orbit is the quotient $\mathbf{SU}(1, 1)/\mathbf{H}$, where  $\mathbf{H}=\left\{\begin{pmatrix}
    \cosh{\rho}& \sinh{\rho}\\
    \sinh{\rho}&\cosh{\rho}
\end{pmatrix}+i\,b \begin{pmatrix}
    1& -1\\
    1&-1
\end{pmatrix}\,,\; \rho, b\in \mathbb{R}\right\}$. It is easy to see that $\mathbf{H}\simeq \mathbb{R}_+\ltimes \mathbb{R}$.} $\mathbb{S}^1$ parametrized by $\theta$. The function $\Phi_\chi(z,w| \theta)$ has no zeros or poles in $z$ and $w$ inside the unit circle, and its only singularities may occur at the boundary. The energy of the above state is given by~(\ref{FormalWavesHyperbolic}) as
\begin{align}
    E=\chi(1-\chi)\,.
\end{align}

We will see shortly that the wave functions~(\ref{Phiwavefunc}) are $\delta$-normalizable for a certain range of $\chi$'s, which signals a continuum spectrum. This range is
\begin{align}\label{lambdarange}
    \chi={1\over 2}+i s\,,\quad\quad s>0\,,
\end{align}
where the parameter $s$ plays the role of the incoming momentum in a scattering state. The corresponding energy range is therefore $E={1\over 4}+s^2$. For the time being, let us take this for granted and study the scalar product\footnote{To simplify notation, we will occasionally suppress the explicit dependence on $z$ and $w$ in the formulas that follow.} $\big(\Phi_{\chi^\prime}(\theta^\prime), \Phi_\chi(\theta)\big)$. We prove in Appendix~\ref{scalprodapp} that considerations of symmetry alone lead to the following unique answer:
\begin{align}\label{scalprodPhi}
    \big(\Phi_{\chi^\prime}(\theta^\prime), \Phi_\chi(\theta)\big)=\upalpha \, \delta\big(s-s^\prime\big) \delta\big(\theta-\theta^\prime\big)
\end{align}
with $\upalpha$ a constant.

If $\Phi_\chi(\theta)$ is an eigenfunction of the Hamiltonian $\widehat{H}_{\mathrm{spin}}$ \eqref{SpinHyperbolicHamiltonian}, then so is the linear combination
\begin{align}\label{flambdaeigenfunc}
f_\chi:=\int\limits_0^{2\pi}\,\dif\theta\,f(\theta)\,\Phi_\chi(\theta)
\end{align}
for $f(\theta)$ an arbitrary function on the circle $\mathbb{S}^1$. The expression~(\ref{scalprodPhi}) then leads to the following form of the scalar product for such general eigenfunctions:
\begin{align}
    (g_{\chi'}, f_{\chi})=\upalpha\,\delta(\chi-\chi')\,\int\limits_0^{2\pi}\,\dif\theta\,\smallthickbar{g}(\theta)\,f(\theta)\,
\end{align}
which is exactly the scalar product for representations of the principal series modeled on functions on a circle $\mathbb{S}^1$ (cf.~\cite{PerelomovBook}).

Apart from the $\Phi_\chi$ eigenfunctions of our holomorphic Hamiltonian \eqref{SpinHyperbolicHamiltonian}, we will be interested in the `physical' eigenfunctions  $\mathcal{F}_\chi$ of the Laplace operator, which
are proportional to the restrictions of $\Phi_\chi$ to~$w=\smallthickbar{z}$:\footnote{This expression is 
 referred to as 
the bulk-to-boundary propagator in the context of the AdS/CFT correspondence~\cite{WittenAdS}. It is also a power of the so-called Poisson kernel \cite[Chapter~V, section~2]{Balazs:1986uj}.}
\begin{align}\label{PoissonKernel}
    \mathcal{F}_{\chi}(z, \smallthickbar{z})=\big(1-|z|^2\big)^{p}\,\Phi_{\chi}\big|_{w=\bar{z}}=\big(1-|z|^2\big)^{\chi}\, \big|1-e^{i\theta}z\big|^{-2\chi} \,,\quad\quad \theta\in \mathbb{S}^1\,.
\end{align}
As explained in Appendix \ref{AppendixFlatRelation}, the extra term $\big(1-|z|^2\big)^{p}$ is required to trivialize the quantum line bundle $\mathscr{O}(-p) \boxtimes \mathscr{O}(-p)$ over $\mathbb{H}_{p}^+ \times \mathbb{H}_{p}^-$, thereby allowing the state to be represented by a scalar function rather than a section.

As we will now review, these wavefunctions are $\delta$-normalizable for the same range~(\ref{lambdarange}) of $\chi$, and by the symmetry considerations elaborated in Appendix~\ref{scalprodapp} their scalar product  differs from~(\ref{scalprodPhi}) only by the  value of $\upalpha$ (an alternative argument is based on Schur's lemma, see  Appendix~\ref{NormAppendix}).

\subsubsection{Normalizability.}\label{HyperbolicContNormalizability}

Before studying the issue of normalizability it will be convenient to slightly smoothen the functions $\Phi_\chi$ and $\mathcal{F}_\chi$, since the normalization of the latter contains an additional angular delta-function. We will thus consider the Fourier zero modes\footnote{One could as well consider higher Fourier coefficients, but this leads only to additional technical complications and does not change the result.}
\begin{align}\label{Tfunc}
&T_{\chi}:= {1\over 2\pi}\int\limits_{0}^{2\pi}\,\dif\theta\,\Phi_\chi(\theta)=(1-z w)^{\chi-p}\,{}_2F_1(\chi, \chi, 1 , zw)\,,\\
&S_{\chi}:= {1\over 2\pi}\int\limits_{0}^{2\pi}\,\dif\theta\,\mathcal{F}_\chi(\theta)=\big(1-|z|^2\big)^{\chi}\,{}_2F_1\big(\chi, \chi, 1 , |z|^2\big)\,.
\end{align}

First let us recall the more familiar situation of the $\mathcal{F}$- and $S$-functions. Here the scalar product is the usual one for $L^2(\mathbb{H})$:
\begin{align}\label{Sscalprod0}
    (S_{\chi^\prime}, S_{\chi})=\int\,\frac{\dif^2z}{\big(1-|z|^2\big)^2}\,\big(1-|z|^2\big)^{\chi+\widebar{\chi}^\prime} {}_2F_1\big(\chi^\prime, \chi^\prime, 1, |z|^2\big) \;{}_2F_1\big(\chi, \chi, 1, |z|^2\big)\,.
\end{align}
The hypergeometric function ${}_2F_1\big(\chi, \chi, 1, w\big)$ is regular for $|w|<1$ and has a singular point at $w=1$, so normalizability depends on the behavior at that point, which corresponds to $|z|^2=1$. At $w=1$ one can expand the hypergeometric function as 
\begin{align}\label{hypergeomexp}
    {}_2F_1\big(\chi, \chi, 1, w\big)=\left(\frac{\Gamma(1-2\chi)}{\Gamma(1-\chi)^2}+\ldots\right)+(1-w)^{1-2\chi}\left(\frac{\Gamma(2\chi-1)}{\Gamma(\chi)^2}+\ldots\right)\,,
\end{align}
where the ellipsis stands for a Taylor expansion in each case. One then easily sees that for $\mathrm{Re}(\chi)>{1\over 2}$ or $\mathrm{Re}(\chi)<{1\over 2}$ (and similarly for $\chi^\prime$) the integral~(\ref{Sscalprod0}) has a power-like divergence. 

The boundary case, when the divergence is formally logarithmic, is when $\mathrm{Re}(\chi)={1\over 2}$, i.e.~$\chi={1\over 2}+is$ with $s\in\mathbb{R}$. In this case, making the change of variables $e^x=1-|z|^2$, one finds the following behavior for $|z|\to 1$ (or $x\to -\infty$):
\begin{align}\label{2F1exp1}
    \big(1-|z|^2\big)^{\chi-1/2} {}_2F_1(\chi, \chi, 1, |z|^2)=\frac{\Gamma(-2is)}{\Gamma\big({1\over 2}-is\big)^2}\,e^{isx}+\frac{\Gamma(2is)}{\Gamma\big({1\over 2}+is\big)^2}\,e^{-isx}+\ldots
\end{align}
The remaining part of the integration measure in the radial direction becomes simply $\int\,dx$, so that the scalar product looks asymptotically at $|z|\to 1$ as a scalar product of plane waves. It turns out that in this case   the wave functions admit a delta-function normalization. The idea is that the asymptotic regions (or here a single region ${x\to -\infty}$) contribute a $\delta$-function to the value of the integral~(\ref{Sscalprod0}), whereas the remaining integral is finite, so that one can write
\begin{align}
    (S_{\chi^\prime}, S_{\chi})=A\, \delta(s-s^\prime)+R(s, s^\prime)\,,
\end{align}
with $A$ a constant and $R(s, s^\prime)$ a smooth function. However, since $S_{\chi}, S_{\chi^\prime}$ are eigenfunctions of a Hermitian Hamiltonian corresponding to different eigenvalues for $s\neq s^\prime$, they are orthogonal, so that $R(s, s^\prime)=0$. Then, by continuity, $R(s, s^\prime)$ vanishes identically, and $(S_{\chi^\prime}, S_{\chi})=A\, \delta(s-s^\prime)$. It is then clear that the constant $A$ may be found from the asymptotic plane wave expansion, such as~(\ref{2F1exp1}). In our case one gets\footnote{We make use of the identity $\Gamma(z)\Gamma(1-z)={\pi \over \sin{\pi z}}$.}
\begin{align}\label{Sscalprodnorm}
(S_{\chi^\prime}, S_{\chi})\sim \frac{\Gamma(2is)\Gamma(-2is)}{\Gamma(1/2+is)^2\Gamma(1/2-is)^2}\delta(s-s^\prime)\sim \frac{(\cosh{\pi s)^2}}{s \sinh 2\pi s} \delta(s-s^\prime)\,.
\end{align}

Next, we pass to the normalization of the $T$-functions~(\ref{Tfunc}). Here the scalar product has the form~(\ref{scalp1p2}):
\begin{align}\label{TF21scalprod}
    (T_{\chi^\prime}, T_{\chi})=\int\,\mathrm{vol}_{p,p}\,(1-z w)^{\chi-p}(1-\smallthickbar{z}\smallthickbar{w})^{\widebar{\chi}^\prime-p} {}_2F_1(\chi^\prime, \chi^\prime, 1, z w) \;{}_2F_1(\chi, \chi, 1, z w)\,.
\end{align}
We study this integral in detail in Appendix~\ref{Tscalprodapp}. 
The idea is to expand the integrand in a power series in $z w$, integrate explicitly over $w$ term by term, and then evaluate the behavior of the remaining series at $|z|\to 1$. Just like in the case of the $S$-functions, this leads to a $\delta$-function normalization for $\chi ={1\over 2}+i s, \chi^\prime={1\over 2}+i s^\prime$ and $s, s^\prime>0$, but this time with a different coefficient:
\begin{align}\label{Tscalprodnorm} 
(T_{\chi^\prime}, T_{\chi}) \sim \frac{(\cosh{\pi s})^2}{s\sinh{2\pi s}}\,\frac{\big[(p-2)!\big]^2}{|\Gamma(p-1/2-is)|^2}\delta(s-s^\prime)\,,
\end{align}
in agreement with \cite[formula (147)]{KitaevSL2R}. Comparing~(\ref{Sscalprodnorm}) and~(\ref{Tscalprodnorm}), we find that in order for the mapping $\mathcal{D}_p^+\otimes \mathcal{D}_p^-\mapsto L^2(\mathbb{H})$ to be an isometry one should map the basis elements of the two spaces as follows:
\begin{align}\label{HyperbolicWavesMap}
    T_\chi \mapsto {(p-2)!\over |\Gamma(p-\chi)|} \,S_\chi\,,\quad \quad \chi={1\over 2}+is\,,
\end{align}
up to a multiplication by a $\chi$-depedent phase, which does not change the values of the scalar products. 

\subsubsection{Flat space limit.} It is instructive to see how the wave functions $\Phi_\chi(z,w|\theta)$ and $\mathcal{F}_\chi(z, \smallthickbar{z}, \theta)$ converge to the flat space wave functions summarized in the table in Appendix~\ref{AppendixFlatRelation} (see also~(\ref{FlatPlaneWaves})), when the radius $R$ of $\mathbb{H}$ tends to infinity. The limit is technically similar to the analogous limit we discussed for the sphere $\mathbb{S}^2$, namely we rescale $(z, w)\to {1\over R}(z, w)$ and set $s=R\,k, p=R^2\lambda$ prior to sending $R\to \infty$. For the wave functions we then get the following limits:
\begin{align}
    &\Phi_\chi\left({z\over R}, {w\over R}\Big|\theta\right)\to \exp\big({\lambda z w+i ke^{i\theta}z+i ke^{-i\theta }w}\big)\,,\\
    &\mathcal{F}_\chi\left({z\over R}, {\smallthickbar{z}\over R}\Big| \theta\right)\to \exp\big({i ke^{i\theta}z+i ke^{-i\theta }\smallthickbar{z}}\big)\,.
\end{align}
When computing scalar products, one should take into account that the variables are rescaled by factors of $R$, so that
\begin{align}
    &\int\limits_{|z|<1}\,{\dif^2z \over \big(1-|z|^2\big)^2}\cdots \mapsto {1\over R^2}\,\int\limits_{\mathbb{C}}\,\dif^2z\,\cdots,\\ &\hspace{-0.5cm}\int\limits_{|z|<1,\,|w|<1}\mathrm{vol}_{p, p}\cdots\mapsto {1\over R^4}\,\int\limits_{\mathbb{C}\times \mathbb{C}}\,\dif^2z\,\dif^2w \,e^{-\lambda |z|^2-\lambda |w|^2}\,\cdots
\end{align}
This is compatible with the scalar products~(\ref{Sscalprodnorm}) and~(\ref{Tscalprodnorm}), since in the large-$R$ limit they behave as follows:
\begin{align}
    &(\mathcal{F}_{\chi^\prime}, \mathcal{F}_{\chi})\sim{1\over R^2}\,{1\over k}\delta(k-k^\prime)\delta(\theta-\theta^\prime)\sim{1\over R^2}\,\delta^{(2)}(k-k^\prime)\,, \\
    &(\Phi_{\chi^\prime}, \Phi_{\chi})\sim{1\over R^4 \lambda}\,e^{{k^2\over \lambda}}\,\delta^{(2)}(k-k^\prime)\,,
\end{align}
where in obtaining the second expression we have expanded the ratio of Gamma-functions in~(\ref{HyperbolicWavesMap}) using Stirling's formula $\log{\Gamma(x)}=x\log{x}-x-{1\over 2}\log(x)+\mathrm{const}+\ldots$ Notice that the exponent in the last line matches precisely the analogous exponent in~(\ref{FlatPlaneWaves}), which is an additional confirmation of the identification~(\ref{HyperbolicWavesMap}).

\section{Discussion}\label{DiscussionSection}

In the present paper we pursued a non-standard approach to the quantization of a point particle moving on a two-dimensional manifold $\mathbb{M}$ of constant curvature (either positive, zero, or negative) and in a homogeneous magnetic field. The idea was to replace the original classical system with an equivalent one, whose phase space is the product of two coadjoint orbits of the symmetry group in question, and then quantize this second system. These orbits are K\"ahler, and one can quantize them in a complex polarization, which leads to the fact that all our wavefunctions are biholomorphic in the two complex coordinates, which we have called $z$ and $w$. To restore the eigenfunctions of the original Hamiltonian it suffices to restrict to $w=\smallthickbar{z}$, which is a Lagrangian submanifold of the original phase space. Our approach makes manifest the fact that the Hilbert space of the original system, which is $L^2(\mathbb{M})$, is a tensor product of two irreducible representations of the symmetry group (possibly in the limit of `large spin', as in the case of $\mathbb{M}=\mathbb{S}^2$). The most interesting consequence of this is a geometric interpretation of the result of Repka~\cite{Repka}, which states that $L^2(\mathbb{H})$ is a tensor product of two representations of the discrete series of $\mathbf{SL}(2,\mathbb{R})$.

We envision several directions for generalizations of our work. First, staying within the realm of two-dimensional target spaces, one might generalize to the case of higher-genus Riemann surfaces, which,  by analogy with flat tori, can be seen as quotients of $\mathbb{H}$ (work in this direction has been done in the context of the quantum Hall effect~\cite{Iengo:1993cs, Klevtsov:2017fwn, Klevtsov:2015nca, Carlos, Hatsugai}). Second, one could generalize to groups of higher rank, such as $\mathbf{SU}(1, n)$, and to higher-dimensional hyperbolic manifolds~\cite{Gitman:1993dh}. A further extension would be to move to non-symmetric complex homogeneous spaces, for example the flag manifolds of $\mathbf{SU}(1, n)$ and other non-compact groups\footnote{In the $\mathbf{SU}(n)$ case sigma models on flag manifolds have been related to Gaudin models and solved in special cases using our method in~\cite{BykovKuzovchikovGaudin}.}. It would as well  be instructive to extend the method for the description of particles on AdS spaces (i.e., to switch to Minkowski signature). 
Finally, our technology is well compatible with supersymmetry~\cite{Bykov:2023uwb, BKK}, and one might construct supersymmetric `spin chains' describing the spectra of the Laplace-Beltrami operator acting on forms, as well as of the Dirac operator.

\vspace{1cm}
\textbf{Acknowledgments.} Sections 1-4 were supported by the Russian Science Foundation grant № 25-72-10177 (\href{https://rscf.ru/en/project/25-72-10177/}{\emph{https://rscf.ru/en/project/25-72-10177/}}). Sections 5-6 were supported by the Moscow Center of Fundamental and Applied Mathematics of Lomonosov Moscow State University under agreement № 075-15-2025-345. We would like to thank  A.~Kuzovchikov and I.~Sechin for useful discussions.

\appendix
\section{Zero magnetic field limit on the plane} \label{zeromagnetic app}
Let us as well discuss the limit of vanishing magnetic field for the plane wavefunctions.
In other words, we want to find a relation between \eqref{FlatPlaneWaves} and \eqref{PhiNstate}.
To this end one should send
\begin{equation}\label{ContLimit}
B\rightarrow 0\,,\qquad n\rightarrow\infty\,,\qquad Bn = E = \mathrm{const}\,.
\end{equation}
A generic state at level $n$ can be seen as a superposition of states $\Phi_n(z,w|\beta)$, introduced in~(\ref{PhiNstate}), with different $\beta$'s, so we shall write
\begin{align}\label{FlatMagneticTransformation}
    \Phi_n[f]:=\int\,\dif^2\beta\, \exp\bigg({-{|\beta|^2\over B}}\bigg) \smallthickbar{\beta}^n f(\smallthickbar{\beta}\,) \,\Phi_n(z,w|\beta)\,.
\end{align}
Notice that we have factored out $\smallthickbar{\beta}^n$; as a result, $f(\smallthickbar{\beta}\,)$ may be a Laurent series starting from $\smallthickbar{\beta}^{-n}$. Upon shifting $\beta \to\beta+z B$ in the above integral, one gets
\begin{align}
    \Phi_n[f]:=\int\,\dif^2\beta\, \exp\bigg({-{|\beta|^2\over B}}\bigg) |\beta|^{2n} f(\smallthickbar{\beta}\,) \,e^{(\lambda+B) z w+\beta w-\widebar{\beta} z }\,.
\end{align}
We may now set $B={E\over n}$ and send $n\to \infty$. In this limit the integral is dominated by the saddle point at $|\beta|^2=E$, so that
\begin{align}\label{fixedenergyexp}
    \Phi_n[f]\sim \int\,\dif^2\beta\, \delta\big(|\beta|^2-E\big) f(\smallthickbar{\beta}\,) \,e^{\lambda z w+\beta w-\widebar{\beta} z }\,.
\end{align}
In other words, we return to the expansion~(\ref{Phithetaintegral})  in plane waves
\begin{align}
    \Phi(z,w) =  \mathrm{exp}\left(\lambda zw+\beta w-{\smallthickbar{\beta}} z\right)\,,\quad\quad |\beta|^2=E\,,\label{LambdazwWaves}
\end{align}
with a fixed absolute value of momentum.

\section{On Hilbert space invariant products}\label{NormAppendix}
As a useful aside, let's consider how we can endow our space of quantum states for the particles on the sphere and hyperbolic plane with a scalar product.

We start with the case of the sphere.
Note that the classical action \eqref{ActionSphere} is invariant under a diagonal $\mathbf{SU}(2)$, which simultaneously rotates the $z$ and $w$ vectors. Thus, the reasonable requirement on the scalar product $(\bullet,\bullet)$ is that it should be invariant under the diagonal $\mathbf{SU}(2)$ action, 
\begin{equation}
\big(g\circ f(z,w),g\circ h(z,w)\big) := \big(f(gz, gw), h(gz,gw)\big) = \big(f(z,w),h(z,w)\big)\,,
\end{equation}
where $g$ is an $\mathbf{SU}(2)$ rotation and the functions are homogeneous of degree $p$ in  $z$ and of degree $p+\mathfrak{q}$ in $w$. Thus, the general admissible form of the scalar product is
\begin{equation}\label{ZWsphereMeasure}
\big(f(z,w),h(z,w)\big) = \int\dif^4 z\,\dif^4 w\,\overline{f(z,w)}h(z,w)\,\mathcal{F}\big(|z|^2,|w|^2\big)\,,
\end{equation}
where $\mathcal{F}\big(|z|^2,|w|^2\big)$ is a suitable measure factor such that the integral converges, for example $\mathcal{F}\big(|z|^2,|w|^2\big) = \exp\big(-|z|^2-|w|^2\big)$.
Using integration by parts, one can also check that, with this measure,
the Hermitian conjugate of 
$A = w^\alpha\frac{\partial}{\partial z^\alpha}$ is $A^\dagger = z^\alpha\frac{\partial}{\partial w^\alpha}$,  
and thus the Hamiltonian \eqref{SphereHamilt} is self-adjoint and positive-definite.

As discussed earlier, the eigenfunctions \eqref{WellDefinedSectionsSphere} transform under irreducible representations of $\mathbf{SU}(2)$. In fact, the number of integrals required to compute the scalar product between such functions can be significantly reduced. A direct consequence of Schur's lemma states that all invariant scalar products on a unitary irreducible representation are proportional up to a constant.
Indeed, suppose we have two invariant products $(\bullet,\bullet)_1$ and $(\bullet,\bullet)_2$. Choose a vector $v$ in the representation space and consider the linear functional $(v,\bullet)_2$. By the Riesz representation theorem, there exists another vector $Av$, for some operator $A$, such that $(v,\bullet)_2 = (Av,\bullet)_1$. Using the invariance of both products, one can show that $A$ commutes with the group action, and therefore, on an irreducible representation, $A$ must be proportional to the identity operator. In other words,
\begin{equation}
(\bullet,\bullet)_1\big|_{\mathrm{irrep}} = \mathrm{const}\cdot(\bullet,\bullet)_2\big|_{\mathrm{irrep}}\,,
\end{equation}
where the constant depends on the specific irreducible representation.

Thus, we can define another invariant product by the formula
\begin{equation}\label{ReducedMeasureSphere}
\big(f(z,w),h(z,w)\big) = \int\dif^4 z\,\overline{f(z,\smallthickbar{z}\varepsilon)}\,h(z,\smallthickbar{z}\varepsilon)\,\tilde{\mathcal{F}}\big(|z|^2\big)\,,
\end{equation}
where we formally restrict the complex coordinate $w$ to the Lagrangian subspace $\big(\mathbb{S}^2\big)_{\mathrm{min}}\simeq\{\smallthickbar{z}\circ w =0\}$, i.e.~we set $w^i = \varepsilon^{ij}\smallthickbar{z}_j$. Here, $\tilde{\mathcal{F}}\big(|z|^2\big)$ is another admissible function that ensures the convergence of the integral. As discussed earlier, on irreducible representations of $\mathbf{SU}(2)$, which correspond to the Hamiltonian eigenfunctions, the reduced measure \eqref{ReducedMeasureSphere} coincides with \eqref{ZWsphereMeasure} up to a constant factor, depending on the specific representation.

We now clarify how to express the scalar product \eqref{ZWsphereMeasure} in inhomogeneous coordinates. We demonstrate this for the $z$ variable; the treatment for $w$ is completely analogous. The required manipulations are as follows:
\begin{align}
&\big(f(z),h(z)\big) = \int\dif^4 z \,\overline{f(z)}h(z)\mathcal{F}\big(|z|^2\big) 
= \int\dif^2 z_1\dif^2 z_2\,\overline{f(z_1,z_2)}\,h(z_1,z_2)\,{\mathcal{F}}\big(|z_1|^2+|z_2|^2\big)=\nonumber\\
&=[z_2\mapsto z_2\,z_1]= \int\dif^2 z_1\dif^2 z_2\,|z_1|^{2+2p}\,\overline{f(z_2)}\,h(z_2)\,{\mathcal{F}}\Big(|z_1|^2\big(1+|z_2|^2\big)\Big)=\nonumber\\
&=\big[z_1\mapsto\big(1+|z_2|^2\big)^{-\frac{1}{2}}z_1\big]=\underbrace{\int \dif^2 z_1\,{\mathcal{F}}\big(|z_1|^2\big)|z_1|^{2+2p}}_{\text{normalization constant}}\int\frac{\dif^2 z_2\,\overline{f(z_2)}\,h(z_2)}{\big(1+|z_2|^2\big)^{2+p}}\,.\label{ReductionHomogeneousSphere}
\end{align}
Thus, we obtain the standard measure on sections of $\mathscr{O}(p)$ written using the inhomogeneous coordinate~$z_2$. 

The case of the hyperbolic plane is largely analogous; let us outline the key differences. Here, the space of states consists of bi-holomorphic functions that are homogeneous of degree $-p$ in $z$ and degree $-p-\mathfrak{q}$ in $w$, defined on the domain $\lVert z\rVert^2 > 0$, $\lVert w\rVert^2 < 0$, where $\lVert \bullet\rVert$ denotes the indefinite complex norm. The counterpart of \eqref{ZWsphereMeasure} is
\begin{equation}
\big(f(z,w),h(z,w)\big) = \int\limits_{\lVert z\rVert^2 > 0}\dif^4 z\int\limits_{\lVert w\rVert^2 < 0}\dif^4 w\,\overline{f(z,w)}h(z,w)\,\mathcal{F}\big(\lVert z\rVert^2,\lVert w\rVert^2\big)\,,
\end{equation}
where $\mathcal{F}\big(\lVert z\rVert^2,\lVert w\rVert^2\big)$ is again an admissible function. 
Alternatively, via a change of variables, we may integrate over the region $\lVert w\rVert^2 > 0$ with the function $\mathcal{F}\big(\lVert z\rVert^2,-\lVert w\rVert^2\big)$.
Owing to the indefinite nature of $\mathcal{F}$, the conjugation rule acquires an extra minus sign compared to the $\mathbb{S}^2$ case,
\begin{equation}
A = w^\alpha\frac{\partial}{\partial z^\alpha}\qquad\Longrightarrow\qquad A^\dagger = -z^\alpha\frac{\partial}{\partial w^\alpha}\,,
\end{equation}
where we omit an irrelevant multiple depending on $\mathcal{F}$.
This can be understood via integration by parts and the identity
\begin{equation}
w^\alpha\frac{\partial}{\partial z^\alpha} \mathcal{F}\Big(|z_1|^2-|z_2|^2,|w_2|-|w_1|^2\Big) = -\smallthickbar{z}_{\,\alpha}\frac{\partial}{\partial \smallthickbar{w}_\alpha}\hat{\mathcal{F}}\Big(|z_1|^2-|z_2|^2,|w_2|-|w_1|^2\Big)\,,
\end{equation}
where we use that we integrate over one ``time-like'' and one ``space-like'' vector. The new function $\hat{\mathcal{F}}$ arises from taking a derivative with respect to the first argument and an antiderivative with respect to the second. The specific form of $\hat{\mathcal{F}}$ is inconsequential because, as noted earlier, scalar products defined with different such functions are equivalent. 

The reduced scalar product ``on the Lagrangian submanifold'' $(\mathbb{H})_{\mathrm{min}}\simeq\{\smallthickbar{z}\eta w = 0\}$ is given by 
\begin{equation}
\big(f(z,w),h(z,w)\big) = \int\limits_{\lVert z\rVert^2 > 0}\dif^4 z\,\overline{f(z,\smallthickbar{z}\eta\varepsilon)}\,h(z,\smallthickbar{z}\eta\varepsilon)\,\tilde{\mathcal{F}}\big(\lVert z\rVert^2\big)\,,
\end{equation}
which is a counterpart of~\eqref{ReducedMeasureSphere}.
In the analogy with \eqref{ReductionHomogeneousSphere}, the transition to the inhomogeneous coordinates can be obtained by a similar change of variables. In a similar notation, the result is
\begin{align}
\big(f(z),h(z)\big) = \underbrace{\int_\mathbb{C} \dif^2 z_1\,{\mathcal{F}}\big(|z_1|^2\big)|z_1|^{2-2p}}_{\text{normalization constant}}\int_{|z_2|<1}\frac{\dif^2 z_2\,\overline{f(z_2)}\,h(z_2)}{\big(1-|z_2|^2\big)^{2-p}}\,.
\end{align}
If we take into account both the $z$ and $w$ variables, we get precisely the formulas (\ref{scalp1p2})-(\ref{volp1p2}).

\section{Relation to standard wave functions}\label{AppendixFlatRelation}
In this paper the wave functions are holomorphic functions of several variables. This is in contrast to the traditional coordinate representation formalism, where states are typically represented by $L^2$-functions.

Let us start from the case of flat space and discuss how the ground state functions~\eqref{Phi0},
\begin{equation}
\Phi_0(z,w)=f(w)\,\exp\big(\lambda zw\big)\,,
\end{equation}
encode the well-known ground states of the Landau problem on the complex plane $(\mathbb{C})_{\mathrm{min}}\simeq\{z=\smallthickbar{w}\}$. Note that their restriction to  $(\mathbb{C})_{\mathrm{min}}$ in the space $\mathscr{B}(\mathbb{C}^2)$ makes no sense because functions in $\mathscr{B}$ should be holomorphic. Instead, we can define the isometric embedding  $\Big(\mathscr{B}(\mathbb{C}^2),e^{-\lambda|z|^2-(\lambda+B)|w|^2}\Big)\hookrightarrow \Big(L^2(\mathbb{C}^2),\dif^2z\, \dif^2 w\Big)$ by 
\begin{equation}
\Phi(z,w)\mapsto \Phi(z,\smallthickbar{z},w,\smallthickbar{w}) = \exp\bigg(-\frac{\lambda}{2}|z|^2-\frac{\lambda+B}{2}|w|^2\bigg)\Phi(z,w)\,.
\end{equation}
Now, in $L^2(\mathbb{C}^2)$  the restriction to $(\mathbb{C})_{\mathrm{min}}$ is well-defined, and 
\begin{equation}\label{MagneticWavesWithB}
\Phi_0(z,w)\mapsto\Phi_0(z,\smallthickbar{z},w,\smallthickbar{w})\Big|_{(\mathbb{C})_{\mathrm{min}}}=f(w)\exp\bigg(-\frac{B}{2}|w|^2\bigg)
\end{equation}
coincides with the standard form of the ground states.

Let us also discuss the case of vanishing magnetic field.
It is well known that the Laplace eigenfunctions on the plane are the plane waves. As in the $B\neq 0$ case, they can be obtained from \eqref{LambdazwWaves} by  restricting the functions
\begin{equation}\label{PlaneStates}
\Phi(z,\smallthickbar{z},w,\smallthickbar{w}) = \mathrm{exp}\left[-\frac{\lambda}{2}|z|^2-\frac{\lambda}{2}|w|^2+\beta w-{\smallthickbar{\beta}} z+\lambda zw\right]
\end{equation}
to the Lagrangian submanifold $(\mathbb{C})_{\mathrm{min}}$. Indeed,
\begin{equation}
\Phi(z,\smallthickbar{z},w,\smallthickbar{w})\Big|_{(\mathbb{C})_{\mathrm{min}}} = e^{2i\,\mathfrak{I}(\beta w)}\,,   
\end{equation}
where by $\mathfrak{I}(\bullet)$ we denote the imaginary part. 

Note that the constructed map \(
\mathscr{B}(\mathbb{C}^2)%L^2(\mathbb{C}^2)
\rightarrow L^2(\mathbb{C})\) is not an isometry for general functions. However, it can be verified that this map merely rescales  Hamiltonian eigenfunctions, such as~\eqref{FlatPlaneWaves}, by constant factors. Therefore, we may apply the restriction map to \((\mathbb{C})_{\mathrm{min}}\) for the eigenfunctions, but should bear in mind that the isometry between Bargmann--Fock spaces and \(L^2\)-spaces can be rather subtle; see \eqref{AnalogOfFourier} or~(\ref{HyperbolicWavesMap}) and the related discussion.

The same logic can be applied to the eigenfunctions on the sphere, though an additional step is required. For simplicity, we set $\mathfrak{q} = 0$. We wish to compare the functions \eqref{Ttransformation} of the form
\begin{equation}\label{SphereTsection}
\Phi_n(z,w) = \big(\varepsilon_{ij}z^iw^j\big)^{p-n}\,\mathsf{T}_{i_1\ldots i_{2n}}z^{i_1}\ldots z^{i_n} w^{i_{n+1}}\ldots w^{i_{2n}}
\end{equation}
with spherical harmonics. First, note that these wavefunctions are sections of ${\mathscr{O}(p)\boxtimes\mathscr{O}(p)}$, whereas spherical harmonics are ordinary functions on the sphere. However, the equivalence between a particle on the sphere and the mechanics on a product of two spheres, as expressed in \eqref{SymplWithSphere}, holds only over the submanifold $\big\{\varepsilon_{ij}z^iw^j \neq 0\big\}$. Over this submanifold, the section $s_0(z,w) = \big(\varepsilon_{ij}z^iw^j\big)^{p}$ of $\mathscr{O}(p)\boxtimes\mathscr{O}(p)$ is nowhere vanishing, so that every section $s(z,w)$ can be represented as $s(z,w) = f(z,w)s_0(z,w)$, where $f(z,w)$ is now an ordinary function. Hence, we can map \eqref{SphereTsection} to the function
\begin{equation}
Y_n(z,w) = \frac{\mathsf{T}_{\alpha_1\ldots \alpha_{2n}}z^{\alpha_1}\ldots z^{\alpha_n} w^{\alpha_{n+1}}\ldots w^{\alpha_{2n}}}{\big(\varepsilon_{\alpha\beta}z^\alpha w^\beta\big)^{n}}\,.
\end{equation}
The final step is to restrict these functions to the Lagrangian sphere $\{\smallthickbar{z}\circ w = 0\}$, which is achieved by imposing $w_\alpha = \varepsilon_{\alpha\beta}\smallthickbar{z}^\beta$. This yields
\begin{equation}
Y_n(z,\smallthickbar{z}) = \frac{\tilde{\mathsf{T}}_{\alpha_1\ldots \alpha_n}^{\beta_1\ldots \beta_n} z^{\alpha_1}\smallthickbar{z}_{\beta_1}\ldots z^{\alpha_n}\smallthickbar{z}_{\beta_n}}{|z|^{2n}}\,,
\end{equation}
where $\tilde{\mathsf{T}}$, obtained by contracting $\mathsf{T}$ with the $\varepsilon$ tensors, is a traceless tensor with respect to the contraction of upper and lower indices. Alternatively, these functions may be seen as  restrictions of harmonic homogeneous polynomials to the sphere (see~\cite{Bykov:2023afs} and Theorem 1 in~\cite{Bykov:2023uwb}), thus providing a connection to the  classical approach to  spherical harmonics.

The case of hyperbolic wave functions is nearly identical to the spherical one: the `physical' eigenfunctions on $(\mathbb{H})_{\mathrm{min}}$ are obtained by dividing by the factor $\varepsilon_{\alpha\beta}z^\alpha w^\beta$ raised to an appropriate power and then restricting to the Lagrangian submanifold~$\smallthickbar{z}\eta w = 0$; see, for example, formula \eqref{PoissonKernel}.

The moral is that the restriction map provides a useful bridge for comparing results in our formalism with the classical ones. 
For the reader's convenience, the relationships among various wave functions are summarized in the table below (all notation can be found in the main text). In the case of the plane we have also relabeled $z \leftrightarrow w$ for uniformity. \renewcommand{\arraystretch}{0.7} 
\begin{table}[H]
\centering
\begin{tabular}{l|l|l|}
\cline{2-3}             & & \\
                        & Biholomorphic section in $z,w$ & $L^2$ function in $z,\smallthickbar{z}$ \\
                        & &    \\ \hline
\multicolumn{1}{|l|}{}                       & & \\
\multicolumn{1}{|l|}{Groundstates on $\mathbb{C}$}                       & $f(z)\exp\big(zw\big)$ & $f(z)\exp\bigg(-\dfrac{B}{2}|z|^2\bigg)$\\
\multicolumn{1}{|l|}{} &  &  \\ \hline
\multicolumn{1}{|l|}{} &  &  \\ 
\multicolumn{1}{|l|}{Excited states on $\mathbb{C}$} & $(wB-\beta)^{n}\,\exp\big(zw+\beta z\big)$ & $(\smallthickbar{z}B-\beta)^{n}\exp\bigg(-\dfrac{B}{2}|z|^2+\beta z\bigg)$ \\ 
\multicolumn{1}{|l|}{} &  &  \\ \hline
\multicolumn{1}{|l|}{} &  &  \\
\multicolumn{1}{|l|}{Plane waves} & $\mathrm{exp}\left(zw+\beta z-{\smallthickbar{\beta}} w\right)$ & $\exp\big({2i\,\mathfrak{I}(\beta z)}\big)$,~~$\beta\in\mathbb{C}$ \\
\multicolumn{1}{|l|}{} &  &  \\ \hline
\multicolumn{1}{|l|}{} &  &  \\
\multicolumn{1}{|l|}{Torus groundstates} & $\exp\bigg(\dfrac{B}{2}z^2+zw\bigg)\vartheta_{n,k}(z|\tau)$  & $\exp\bigg(\dfrac{B}{2}z^2-\dfrac{B}{2}|z|^2\bigg)\vartheta_{n,k}(z|\tau)$ \\
\multicolumn{1}{|l|}{} &  &  \\ \hline
\multicolumn{1}{|l|}{} &  &  \\
\multicolumn{1}{|l|}{Spherical states} & $\big(\varepsilon_{\alpha\beta}z^\alpha w^\beta\big)^{p-n}\big(z\circ\upgamma\big)^n\big(w\circ\upgamma\big)^{n}$ & 
$\dfrac{\big(z\circ\upgamma\big)^n\big(\smallthickbar{z}\varepsilon\circ\upgamma\big)^n}{|z|^{2n}}$,~~$\upgamma\in\mathbb{C}^2/\{0\}$
\\
\multicolumn{1}{|l|}{} &  &  \\ \hline
\multicolumn{1}{|l|}{} &  &  \\
\multicolumn{1}{|l|}{Hyperbolic states} & $\dfrac{\big(\varepsilon_{\alpha\beta}z^\alpha w^\beta\big)^{\frac{1}{2}+is-p}}{\big(z\circ\upgamma\big)^{\frac{1}{2}+is}\big(w\circ\upgamma\big)^{\frac{1}{2}+is}}$, \,$\lVert\upgamma\rVert = 0$  & $\dfrac{\big(1-|z|^2\big)^{\frac{1}{2}+is}}{\big|1-e^{i\theta}z\big|^{1+2is}}$,~~$\theta\in\mathbb{S}^1$,~~$|z|<1$   \\
\multicolumn{1}{|l|}{} &  &  \\ \hline
\end{tabular}
\end{table}

\section{Scalar products of wave functions on $\mathbb{H}$}\label{scalprodapp}

Here we come to the proof of~(\ref{scalprodPhi}). Let us denote the scalar product as 
\begin{align}
    \big(\Phi_{\chi^\prime}(\theta^\prime), \Phi_\chi(\theta)\big):= F_{\chi, \chi^\prime}\big(e^{i\theta}, e^{i\theta^\prime}\big)\,.
\end{align}
Since the angular variables transform under $\mathbf{SU}(1,1)$, this places severe restrictions on~$F$. To find the corresponding relation we perform an $\mathbf{SU}(1,1)$ transformation \big(by the group element~(\ref{SU11mat})\big) in the integral defining the scalar product. More exactly, we map
\begin{align}
    w\mapsto {\alpha w
    +\beta\over \smallthickbar{\beta} w+\smallthickbar{\alpha}}, \qquad\qquad z\mapsto {\smallthickbar{\alpha} z+\smallthickbar{\beta}\over \beta z+\alpha}\,.
\end{align}
Setting for simplicity $p_1=p_2=p$, we find the relation
\begin{align}\label{Ftransformation}
    &F_{\chi, \chi^\prime}\Big(g\circ e^{i\theta}, g\circ e^{i\theta^\prime}\Big)=\big|\alpha-e^{i\theta} \smallthickbar{\beta}\big|^{2\chi}\,\big|\alpha-e^{i\theta^\prime}\smallthickbar{\beta}\big|^{2\widebar{\chi}^\prime}F_{\chi, \chi^\prime}\Big( e^{i\theta}, e^{i\theta^\prime}\Big),\\
    &\textrm{where} \quad g\circ e^{i\theta}=\frac{\smallthickbar{\alpha} e^{i\theta}-\beta}{\alpha-e^{i\theta}\smallthickbar{\beta}\,}\,.
\end{align}
There are two options:
\begin{itemize}
\item 
If $F_{\chi, \chi^\prime}\big(e^{i\theta}, e^{i\theta^\prime}\big)\neq 0$ for $\theta\neq \theta'$, let us set $e^{i\theta}=1, e^{i\theta'}=-1$, then the above relation gives
\begin{align}
    &F_{\chi, \chi^\prime}\bigg(\frac{\smallthickbar{\alpha}-\beta}{\alpha-\smallthickbar{\beta}}, -\frac{\bar{\alpha}+\beta}{\alpha+\smallthickbar{\beta}}\bigg)=\big|\alpha-e^{i\theta} \smallthickbar{\beta}\big|^{2\chi} \,\big|\alpha-e^{i\theta^\prime}\smallthickbar{\beta}\big|^{2\widebar{\chi}^\prime}F_{\chi, \chi^\prime}( 1, -1),
\end{align}
First, setting $\alpha$ and $\beta$ real, and using that $F_{\chi, \chi^\prime}( 1, -1)\neq 0$, one finds 
\begin{equation}
1=|\alpha- \beta|^{2\chi} |\alpha+\beta|^{2\widebar{\chi}^\prime}
\end{equation}
leading to $\smallthickbar{\chi}^\prime=\chi$. In this case the above relation is easily solved to give
\begin{align}
    F_{\chi, \chi^\prime}\big(e^{i\theta}, e^{i\theta^\prime}\big)=\frac{\upalpha}{\big|\sin{{\theta-\theta^\prime\over 2}}\big|^{2\chi}}\delta\big(\chi-\chi^\prime\big),\quad\quad \upalpha=F_{\chi, \chi^\prime}( 1, -1)\,.
\end{align}
This corresponds to the complementary series, with the range of $0<\chi, \chi^\prime<{1\over 2}$. 
\item The second option is when $F_{\chi, \chi^\prime}\big(e^{i\theta}, e^{i\theta^\prime}\big) \equiv 0$ for all $\theta\neq \theta'$, in which case a natural ansatz is $F_{\chi, \chi^\prime}\big(e^{i\theta}, e^{i\theta^\prime}\big)=\upalpha\, \delta\big(\theta-\theta^\prime\big) \varphi(\theta)=\upalpha \,\delta\big(e^{i\theta}-e^{i \theta^\prime})\widehat{\varphi}(\theta)$, where $ \widehat{\varphi}(\theta)=ie^{i\theta} \varphi(\theta)$. We then find the following relation for $\varphi(\theta)$:
\begin{align}
    \varphi\left(\frac{\smallthickbar{\alpha} e^{i\theta}-\beta}{\alpha-e^{i\theta}\smallthickbar{\beta}}\right)=\big|\alpha-e^{i\theta} \smallthickbar{\beta}\big|^{2(\chi+\widebar{\chi}^\prime-1)} \varphi(\theta)\,.
\end{align}
Setting first $\beta=0$, one easily finds that $\varphi(\theta)=\mathrm{const.}$  This obviously leads to $\chi+\smallthickbar{\chi}^\prime-1=0$, which by the unitarity constraint implies that $\chi={1\over 2}+i s$ for~$s>0$.
\end{itemize}

By virtue of this calculation, any two scalar products with the transformation property~(\ref{Ftransformation}) and with the same range of $\chi$ are proportional to each other. As above, here we have the following two scalar products in mind:
\begin{align}
    &\big(\Phi_{\chi^\prime}(z, w), \Phi_{\chi}(z, w)\big)=\int\,\frac{\dif^2z\,\dif^2w}{(1-|z|^2)^{2-p}(1-|w|^2)^{2-p}}\,\smallthickbar{\Phi}_{\chi^\prime}\,\Phi_{\chi}\,,\\
    &\big(\mathcal{F}_{\chi'}(z, \bar{z}), \mathcal{F}_{\chi}(z, \bar{z})\big)=\int\,\frac{\dif^2z}{(1-|z|^2)^2}\,\smallthickbar{\mathcal{F}}_{\chi^\prime} \mathcal{F}_{\chi}\,,
\end{align}
where $\Phi_{\chi}$ are given by~(\ref{Phiwavefunc}) and $\mathcal{F}_{\chi}$ are proportional to their restrictions to $w=\smallthickbar{z}$, i.e.~$\mathcal{F}_{\chi}(z, \smallthickbar{z})=\big(1-|z|^2\big)^{p}\Phi_{\chi}\big|_{w=\bar{z}}$. As shown in Section \ref{HyperbolicContNormalizability}, for both scalar products the wave functions are delta-normalizable over the same range of $\chi$ corresponding to the principal series. Consequently, under the map $\Phi \mapsto \mathcal{F}$ the two scalar products are proportional to each other, with a proportionality factor that generally depends on $p$ and $\chi$. For the zeroth Fourier harmonic this coefficient is given by \eqref{HyperbolicWavesMap}.

\section{Scalar product of the $T$-eigenfunctions}\label{Tscalprodapp}

In this Appendix we study the scalar product  of the $T$-functions:
\begin{align}
    (T_{\chi^\prime}, T_{\chi})=\int\,\mathrm{vol}_{p,p}\,(1-z w)^{\chi-p}(1-\smallthickbar{z}\smallthickbar{w})^{\widebar{\chi}^\prime-p} {}_2F_1(\chi^\prime, \chi^\prime, 1, z w) {}_2F_1(\chi, \chi, 1, z w)\,.
\end{align}
Our goal is to show that the $T$-functions are $\delta$-normalizable for the appropriate range of $\chi$~(\ref{lambdarange}).

We will follow the strategy proposed in~\cite{KitaevSL2R}, that is we expand the integrand in a power series in $z w$, integrate explicitly over $w$ term by term, and then evaluate the behavior of the remaining series at $|z|\to 1$. Since the radius of convergence of the series is $R=1$, the asymptotic will be determined by the decay rate of the series coefficients at $n\to \infty$. One may write
\begin{align}
    (1-z w)^{\chi-p} {}_2F_1(\chi, \chi, 1, z w)=\sum\limits_n\,f_n(\chi)\,(zw)^n\,,
\end{align}
then the scalar product is given by the formula
\begin{align}\label{Tscalprod}
    (T_{\chi^\prime}, T_{\chi})=\pi \int\,\frac{\dif^2z}{\big(1-|z|^2\big)^{2-p}}\,\sum\limits_{n=0}^\infty\,\frac{\Gamma(n+1)\Gamma(p-1)}{\Gamma(n+p)}\widebar{f_n(\chi^\prime)} f_n(\chi)\, |z|^{2n}\,.
\end{align}
\begin{figure}
    \centering
\includegraphics[width=0.4\linewidth]{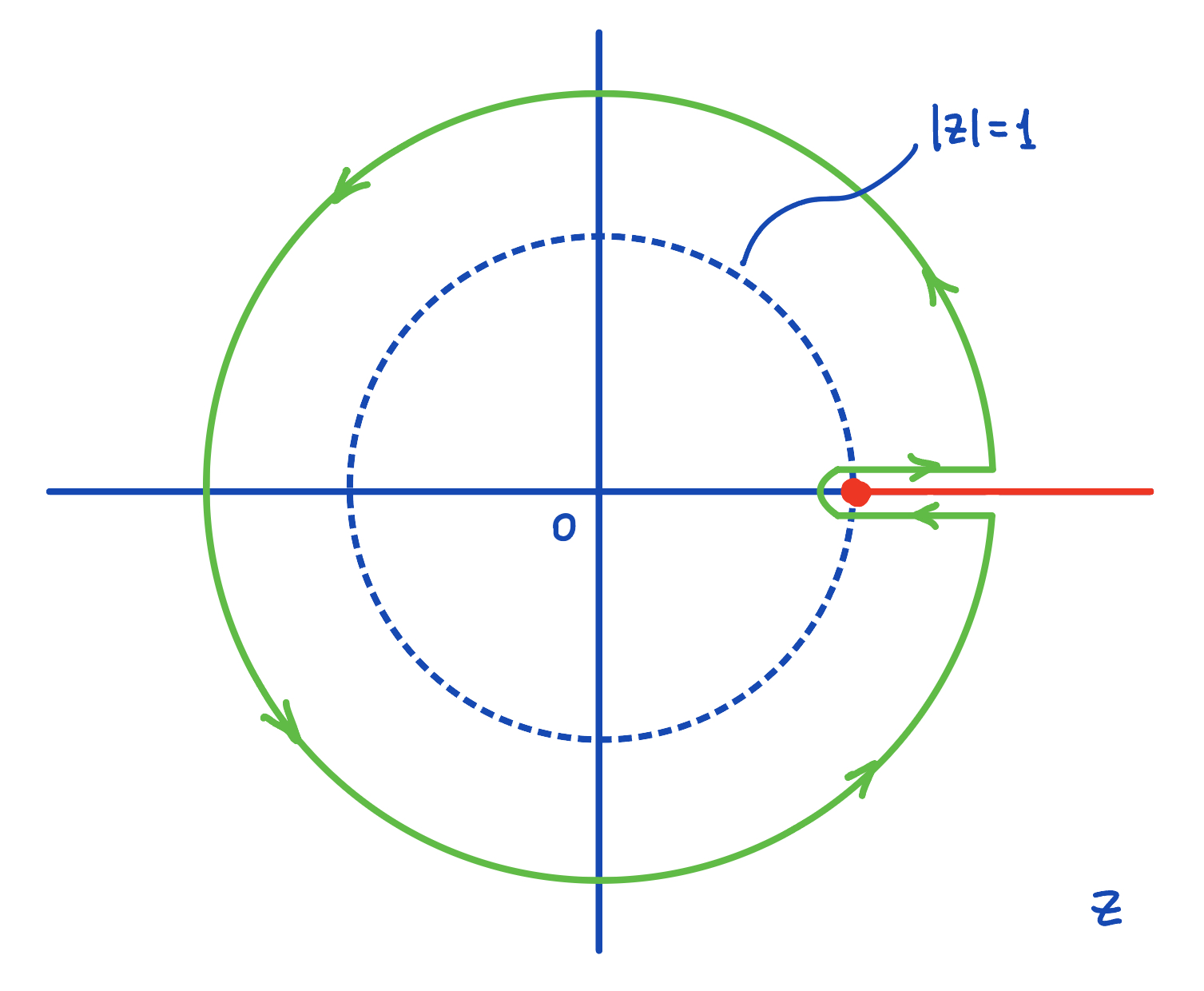}
    \caption{The deformed integration contour $\mathcal{C}$ is shown in green. The radius of the green circle is eventually sent to infinity. The cut $[1, \infty)$ is shown in red.}
    \label{intcontourpic}
\end{figure}

What remains is to find the behavior of the coefficients $f_n(\chi)$ at $n\to \infty$. One may express them as follows:
\begin{align}
    f_n(\chi)={1\over 2\pi i}\oint_{\mathcal{C}}\,{\dif z \over z^{n+1}}(1-z)^{\chi-p} {}_2F_1(\chi, \chi, 1, z)\,.
\end{align}
One may extend the contour to infinity, as shown in Fig.~\ref{intcontourpic}. The integral over the large arc is zero, and one is left with the integral around the positive semi-axis. 
To obtain the leading behavior at $n\to\infty$, one changes variables $z=\exp\big({t\over n}\big)$ and expands in~$n^{-1}$. Thus, we are led to consider the asymptotic behavior of the integrand around $z=1$, which follows from (\ref{hypergeomexp}): 
\begin{align}
    (1-z)^{\chi-p} {}_2F_1(\chi, \chi, 1, z)=\frac{\Gamma(-2is)}{\Gamma\big({1\over 2}-is\big)^2}\,(1-z)^{\frac{1}{2}-p+is}+\frac{\Gamma(2is)}{\Gamma\big({1\over 2}+is\big)^2}\,(1-z)^{\frac{1}{2}-p-is}+\ldots
\end{align}
As a result, we find the following behavior of the  $f_n$ coefficients at large $n$:
\begin{align}
    f_n(\chi)={\cosh{(\pi s)}\over 2\pi i}\Bigg(\frac{\Gamma(-2is)}{\Gamma\big({1\over 2}-is\big)^2}\,{\Gamma\big(\frac{3}{2}-p+is\big)\over n^{\frac{3}{2}-p+is}}+\frac{\Gamma(2is)}{\Gamma\big({1\over 2}+is\big)^2}\,{\Gamma\big(\frac{3}{2}-p-is\big)\over n^{\frac{3}{2}-p-is}}\Bigg)+\ldots
\end{align}
Approximating ${\Gamma(n+1)\over \Gamma(n+p)}\simeq n^{1-p}$ in~(\ref{Tscalprod}), we get the following expression for the scalar product:
\begin{align}
    (T_{\chi^\prime}, T_{\chi})\sim\pi (p-2)!\,\int\,\frac{\dif^2z}{(1-|z|^2)^{2-p}} \sum\limits_{n=1}^\infty\,{\widebar{f_n(\chi^\prime)} f_n(\chi)\over n^{p-1}}\, |z|^{2n}\,.
\end{align}
Clearly this will involve four terms, each schematically of the form 
\begin{align}
\sum\limits_{n=1}^\infty {|z|^{2n}\over n^r}=\mathrm{Li}_r\big(|z|^2\big)\simeq \big(1-|z|^2\big)^{r-1}\,\Gamma(1-r) +\ldots \quad \textrm{for}\quad |z|\to 1\,.     
\end{align}
The relevant exponent is $r=2-p\pm i(s\pm s^\prime)$ in all cases. Since we assume $s, s^\prime>0$, there are only two terms contributing to $\delta(s-s^\prime)$. Collecting all coefficients, we get
\begin{align} \nonumber
    (T_{\chi^\prime}, T_{\chi})&\sim [(p-2)!]^2 (\cosh{\pi s})^2\,\frac{\Gamma(2is)\Gamma(-2is)\Gamma\big(\frac{3}{2}-p+is\big)\Gamma\big(\frac{3}{2}-p-is\big)}{\Gamma\big(\frac{1}{2}+is\big)^2 \Gamma\big(\frac{1}{2}-is\big)^2}\,\delta(s-s^\prime)\sim \\  
    &\sim \frac{(\cosh{\pi s})^2}{s\sinh{2\pi s}}\,\frac{\big[(p-2)!\big]^2}{|\Gamma(p-1/2-is)|^2}\,\delta(s-s^\prime)\,,
\end{align}
in agreement with \cite[formula (147)]{KitaevSL2R}.

\appendix

\printbibliography
\end{document}